\documentclass[11pt]{article}
\usepackage{amssymb,amsmath,amsfonts}
\usepackage{graphicx}
\usepackage{graphics}
\usepackage{eepic,epsfig}

1.60cm \makeatletter \@addtoreset{equation}{section}

\makeatother

\begin{document}

\title{Casimir densities from coexisting vacua}
\author{S. Bellucci$^{1}$\thanks{%
E-mail: bellucci@lnf.infn.it },\, A. A. Saharian$^{2}$ \thanks{%
E-mail: saharian@ysu.am},\, A. H. Yeranyan$^{3,1,2}$\thanks{%
E-mail: ayeran@lnf.infn.it} \\
\\
\textit{$^1$ INFN, Laboratori Nazionali di Frascati,}\\
\textit{Via Enrico Fermi 40, 00044 Frascati, Italy} \vspace{0.3cm}\\
\textit{$^2$Department of Physics, Yerevan State University,}\\
\textit{1 Alex Manoogian Street, 0025 Yerevan, Armenia}\vspace{0.3cm}\\
\textit{$^3$Museo Storico della Fisica e Centro Studi e Ricerche Enrico
Fermi,}\\
\textit{Via Panisperna 89A, 00184, Roma, Italy}}
\maketitle

\begin{abstract}
Wightman function, the vacuum expectation values (VEVs) of the
field squared and the energy-momentum tensor are investigated for
a massive scalar field with general curvature coupling in a
spherically symmetric static background geometry described by two
distinct metric tensors inside and outside a spherical boundary.
The exterior and interior geometries can correspond to different
vacuum states of the same theory. In the region outside the
sphere, the contributions in the VEVs, induced by the interior
geometry, are explicitly separated. For the special case of the
Minkowskian exterior geometry, the asymptotics of the VEVs near
the boundary and at large distances are discussed in detail. In
particular, it has been shown that the divergences on the boundary
are weaker than in the problem of a spherical boundary in
Minkowski spacetime with Dirichlet or Neumann boundary conditions.
As an application of general results, dS and AdS spaces are
considered as examples of the interior geometry. For AdS interior
there are no bound states. In the case of dS geometry and for
nonminimally coupled fields, bound states appear for a radius of
the separating boundary sufficiently close to the dS horizon.
Starting from a critical value of the radius the Minkowskian
vacuum in the exterior region becomes unstable. For small values
of the AdS curvature radius, to the leading order, the VEVs in the
exterior region coincide with those for a spherical boundary in
Minkowski spacetime with Dirichlet boundary condition. The
exceptions are the cases of minimal and conformal couplings: for a
minimal coupling the VEVs are reduced to the case with Neumann
boundary condition, whereas for a conformally coupled field there
is no reduction to Dirichlet or Neumann results.
\end{abstract}

\bigskip


\bigskip

\section{Introduction}

In many physical problems, the model is formulated in backgrounds having
boundaries on which the dynamical variables obey prescribed boundary
conditions. The boundaries can have different physical origins, like
interfaces between two media with different electromagnetic properties in
condensed matter physics, horizons in gravitational physics, domain walls of
various physical nature in the theory of phase transitions and critical
phenomena, branes in string theory and in higher-dimensional cosmologies. In
quantum field theory, the imposition of boundary conditions on a field
operator gives rise to modifications of the spectrum for the vacuum
fluctuations of a quantum field and, as a result, to the change of physical
characteristics of the vacuum state, such as the energy density and vacuum
stresses. As a consequence of this, vacuum forces arise acting on
constraining boundaries. This is the familiar Casimir effect, first
predicted for the electromagnetic field by Casimir in 1948 \cite{Casi48}.
This effect can have important implications on all scales, from subnuclear
to cosmological, and it has been investigated for various types of bulk and
boundary geometries (for reviews see \cite{Most97}-\cite{Casi11}). The
features of the Casimir forces depend on the nature of a quantum field, on
the type of the spacetime manifold, on the geometry of boundaries, and on
the specific boundary conditions imposed on the field. The explicit
dependence can be found for highly symmetric geometries only.

In consideration of the Casimir effect, usually, the boundaries separate the
regions with different electromagnetic properties (for example, media with
different dielectric permittivities). Another type of effect related to the
Casimir physics arises in a class of models with boundaries separating the
spatial regions with different gravitational backgrounds. It can be referred
to as gravitationally induced Casimir effect. The different gravitational
backgrounds on both sides of the separating boundary can correspond to
different vacuum states of the same theory. For example, one can consider a
bubble of a false vacuum embedded in true vacuum or vice versa. Simple
examples of vacuum bubbles are de Sitter (dS) and anti-de Sitter (AdS)
spacetimes embedded in the Minkowski spacetime. In these examples, a
physical boundary separates two regions with different values of the
cosmological constant. It serves as a thin-wall approximation of a domain
wall interpolating between two coexisting vacua (for a discussion see \cite%
{Klin05}).

In a configuration with coexisting gravitational backgrounds, the
geometry of one region affects the properties of the quantum
vacuum in the other region. Previously, we have considered several
examples of this type of vacuum polarization. In \cite{Beze06},
the Casimir densities are investigated for a scalar field in the
geometry of a cosmic string for a core with finite support. In the
corresponding model, the cylindrical boundary separates two
different background geometries: the spacetime outside the
boundary is described by the idealized cosmic string geometry with
a planar angle deficit and for the interior geometry a general
cylindrically symmetric static model is employed. Two specific
models of the core have been considered: the 'ballpoint pen' model
\cite{Hisk85,Gott85}, with a constant curvature interior metric,
and the 'flower pot' model \cite{Alle90} with an interior
Minkowskian spacetime. Similar problems
for the exterior geometry of a global monopole are discussed in \cite%
{Beze06b} and \cite{Beze07} for scalar and fermionic fields, respectively.
In the corresponding models the boundary separating different spatial
geometries is a sphere. The model with a sphere as a boundary and with an
exterior dS metric, described in planar inflationary coordinates, has been
considered in \cite{Milt12}. The vacuum expectation values of the field
squared and the energy-momentum tensor induced by a $Z_{2}$-symmetric brane
with finite thickness located on AdS background are evaluated in \cite%
{Saha07,Saha08} for a massive scalar field. The general case of a static
plane symmetric interior structure for the brane is considered, and the
exterior AdS geometry is described in Poincar\'{e} coordinates. In the
corresponding problem the separating boundaries are plane symmetric.

In the present paper, we consider the vacuum densities for a
massive scalar field with a general curvature coupling parameter
in a spherically symmetric static geometry described by two
distinct metric tensors inside and outside a spherical boundary.
In addition, the presence of a surface energy-momentum tensor
located on the separating boundary is assumed. Among the most
important characteristics of the quantum vacuum are the
expectation values of the field squared and the energy-momentum
tensor. Although the corresponding operators are local, due to the
global nature of the vacuum state, they carry an important
information about the global properties of the bulk. Moreover, in
addition to describing the physical structure of the quantum field
at a given point, the vacuum expectation value (VEV) of the
energy-momentum tensor acts as a source of gravity in the
quasiclassical Einstein equations. Consequently, it plays a
crucial role in modelling a self-consistent dynamics of the
background spacetime. For the evaluation of the VEVs we first
construct the positive frequency Wightman function by the direct
summation over a complete set of scalar modes. This function also
determines the excitation probability of a Unruh-DeWitt detector
(see, for instance, \cite{Birr82}). The quantum effects induced by
distinct geometries in the exterior and interior regions should be
taken into account, in particular, in discussions of the dynamics
of vacuum bubbles during the phase transitions in the early
Universe.

The organization of the paper is as follows. In the next section we describe
the background spacetime under consideration and the matching conditions on
a spherical boundary separating the interior and exterior geometries. A
complete set of normalized mode functions for a scalar field with a general
curvature coupling parameter is constructed in Section \ref{sec:Modes}. By
using the mode functions, in Section \ref{sec:WF} we evaluate the positive
frequency Wightman function for the general case of static spherically
symmetric interior and exterior geometries. This function is presented in
the form where the contribution induced by the interior geometry is
explicitly separated. A special case of the exterior Minkowskian background
is considered in Section \ref{sec:ExtMink}. Explicit expressions for the
VEVs of the field squared and of the energy-momentum tensor are provided and
their behavior in asymptotic regions of the parameters is investigated. As
an application of general results, in Section \ref{sec:dS}, two special
cases of the interior geometry are discussed corresponding to maximally
symmetric spaces with positive and negative cosmological constants (dS and
AdS spaces). Section \ref{sec:Conc} summarizes the main results of the
paper. In Appendix \ref{sec:Append}, the coefficient in the asymptotic
expansion of the logarithmic derivative of the hypergeometric function is
determined, which is used for the evaluation of the leading terms in the
asymptotic expansions of the VEVs near the boundary for the cases of the
interior dS and AdS spaces.

\section{Background geometry}

\label{sec:Geom}

Consider a $(D+1)$-dimensional spherically symmetric static spacetime
described by two distinct metric tensors inside and outside of a spherical
boundary of coordinate radius $r=a$. In the interior region, $r<a$, the
spacetime geometry is regular with the line element%
\begin{equation}
ds_{i}^{2}=e^{2u_{i}(r)}dt^{2}-e^{2v_{i}(r)}dr^{2}-e^{2w_{i}(r)}d\Omega
_{D-1}^{2},  \label{metricinside}
\end{equation}%
where $d\Omega _{D-1}^{2}$ is the line element on a $(D-1)$-dimensional
sphere with a unit radius. The corresponding hyperspherical angular
coordinates will be denoted by $(\vartheta ,\phi )=(\theta _{1},\ldots
,\theta _{n},\phi )$, where $n=D-2$, $0\leqslant \theta _{k}\leqslant \pi $,
$k=1,\ldots ,n$, and $0\leqslant \phi \leqslant 2\pi $. The value of the
radial coordinate $r$ corresponding to the center of the configuration will
be denoted by $r_{c}$. Of course, we could rescale the radial coordinate in
order to have $r=0$ for the center, but for the further discussion it is
convenient to keep $r_{c}$ general. Introducing a new coordinate%
\begin{equation}
\overline{r}=e^{w_{i}(r)},  \label{rbar}
\end{equation}%
with the center at $\bar{r}=0$, the angular components of the metric tensor
coincide with the corresponding components in the Minkowski spacetime
described in the standard hyperspherical coordinates.

In the exterior region, $r>a$, the geometry has a similar structure with
different radial functions:%
\begin{equation}
ds_{e}^{2}=e^{2u_{e}(r)}dt^{2}-e^{2v_{e}(r)}dr^{2}-e^{2w_{e}(r)}d\Omega
_{D-1}^{2}.  \label{metricoutside}
\end{equation}%
The metric tensor is continuous at the separating boundary $r=a$:%
\begin{equation}
u_{i}(a)=u_{e}(a),\;v_{i}(a)=v_{e}(a),\;w_{i}(a)=w_{e}(a).
\label{metricCont}
\end{equation}%
Although the scheme described below can be generalized for metric tensors
with horizons, for the sake of simplicity we will assume that if the line
elements (\ref{metricinside}) and (\ref{metricoutside}) have horizons at $%
r_{Hi}$ and $r_{He}$, respectively, then $r_{He}<a<r_{Hi}$. This means that
the combined geometry contains no horizons.

The Ricci tensors for the interior and exterior geometries are diagonal with
the mixed components (no summation over $l=2,3,\ldots ,D$)%
\begin{eqnarray}
R_{(j)0}^{0} &=&-e^{-2v_{j}}\left[ u_{j}^{\prime \prime }+u_{j}^{\prime
2}-u_{j}^{\prime }v_{j}^{\prime }+(n+1)u_{j}^{\prime }w_{j}^{\prime }\right]
,  \notag \\
R_{(j)1}^{1} &=&-e^{-2v_{j}}\left[ u_{j}^{\prime \prime }+u_{j}^{\prime
2}-u_{j}^{\prime }v_{j}^{\prime }+(n+1)\left( w_{j}^{\prime \prime
}+w_{j}^{\prime 2}-w_{j}^{\prime }v_{j}^{\prime }\right) \right] ,
\label{Riccin} \\
R_{(j)l}^{l} &=&-e^{-2v_{j}}\left( w_{j}^{\prime \prime }+w_{j}^{\prime
2}+w_{j}^{\prime }u_{j}^{\prime }-w_{j}^{\prime }v_{j}^{\prime
}+nw_{j}^{\prime 2}\right) +ne^{-2w_{j}},  \notag
\end{eqnarray}%
where $j=i$ and $j=e$ for the interior and exterior regions respectively and
the prime means the derivative with respect to the radial coordinate $r$ (we
adopt the convention of Ref. \cite{Birr82} for the curvature tensor). For
the corresponding Ricci scalars we get the expression%
\begin{eqnarray}
R_{(j)} &=&-2e^{-2v_{j}}\left[ u_{j}^{\prime \prime }+u_{j}^{\prime
2}-u_{j}^{\prime }v_{j}^{\prime }+n(n+1)w_{j}^{\prime 2}/2\right.  \notag \\
&&\left. +(n+1)\left( w_{j}^{\prime \prime }+w_{j}^{\prime 2}+w_{j}^{\prime
}u_{j}^{\prime }-w_{j}^{\prime }v_{j}^{\prime }\right) \right]
+n(n+1)e^{-2w_{j}}.  \label{Richscin}
\end{eqnarray}%
The energy-momentum tensors generating the line elements (\ref{metricinside}%
) and (\ref{metricoutside}) are found from the corresponding Einstein
equations.

In general, we assume the presence of an infinitely thin spherical shell at $%
r=a$, having a surface energy-momentum tensor $\tau _{i}^{k}$ with nonzero
components $\tau _{0}^{0}$ and $\tau _{2}^{2}=\cdots =\tau _{D}^{D}$. Let $%
n^{i}$, $n_{i}n^{i}=-1$, be the normal to the shell which points into the
bulk on both sides. For the interior ($j=i$) and exterior ($j=e$) regions
one has $n_{i}^{(j)}=\delta _{(j)}\delta _{i}^{1}e^{v_{j}(r)}$ with $\delta
_{(i)}=1$ and $\delta _{(e)}=-1$. We denote by $h_{(j)ik}$ the induced
metric on the shell, $h_{(j)ik}=g_{(j)ik}+n_{i}^{(j)}n_{k}^{(j)}$, and $%
K_{(j)ik}=h_{(j)i}^{l}h_{(j)k}^{r}\nabla _{l}n_{r}^{(j)}$ is the extrinsic
curvature. In the geometry under consideration, for the non-zero components
of the latter we obtain%
\begin{eqnarray}
K_{(j)0}^{0} &=&-\delta _{(j)}u_{j}^{\prime }(r)e^{-v_{j}(r)},  \notag \\
K_{(j)l}^{k} &=&-\delta _{(j)}\delta _{l}^{k}w_{j}^{\prime
}(r)e^{-v_{j}(r)},\;r=a-\delta _{(j)}0,  \label{exttensor}
\end{eqnarray}%
with $l=2,3,\ldots ,D$.

From the Israel matching conditions on the sphere $r=a$ one has%
\begin{equation}
\sum_{j=i,e}(K_{(j)ik}-K_{(j)}h_{(j)ik})=8\pi G\tau _{ik},  \label{matchcond}
\end{equation}%
where $G$ is the gravitational constant and $K_{(j)}=K_{(j)i}^{i}$ is the
trace of the extrinsic curvature tensor. From these conditions, by taking
into account (\ref{exttensor}), we find (no summation over $i=2,3,\ldots ,D$%
):%
\begin{eqnarray}
\sum_{j=i,e}\delta _{(j)}u_{j}^{\prime }(a-\delta _{(j)}0) &=&8\pi
Ge^{v_{e}(a)}\left( \tau _{i}^{i}-\frac{D-2}{D-1}\tau _{0}^{0}\right) ,
\notag \\
\sum_{j=i,e}\delta _{(j)}w_{j}^{\prime }(a-\delta _{(j)}0) &=&\frac{8\pi G}{%
D-1}e^{v_{e}(a)}\tau _{0}^{0},  \label{matchcond2}
\end{eqnarray}%
where $f^{\prime }(a\pm 0)$ is understood as the limit $\lim_{r\rightarrow
a\pm 0}f^{\prime }(r)$. Note that from (\ref{matchcond2}) the relation%
\begin{equation}
\sum_{j=i,e}\delta _{(j)}\left[ u_{j}^{\prime }(a-\delta
_{(j)}0)+(D-1)w_{j}^{\prime }(a-\delta _{(j)}0)\right] =\frac{8\pi G}{D-1}%
e^{v_{e}(a)}\tau ,  \label{Tracesurf}
\end{equation}%
is obtained for the trace $\tau =\tau _{0}^{0}+\sum_{i=2}^{D}\tau _{i}^{i}$
of the surface energy-momentum tensor. For given interior and exterior
geometries, the relations (\ref{matchcond2}) determine the surface
energy-momentum tensor needed for the matching of these geometries.

\section{Mode functions for a scalar field}

\label{sec:Modes}

\subsection{Modes of continuous spectrum}

Having described the background geometry, now we turn to the field content.
We will consider a scalar field $\varphi (x)$ with curvature coupling
parameter $\xi $ on background described by (\ref{metricinside}) and (\ref%
{metricoutside}). The corresponding field equation reads
\begin{equation}
\left( \nabla _{\mu }\nabla ^{\mu }+m^{2}+\xi R\right) \varphi =0,
\label{FieldEq}
\end{equation}%
where $\nabla _{\mu }$ is the covariant derivative operator. The most
important special cases of the curvature coupling parameter $\xi =0$ and $%
\xi =\xi _{D}=(D-1)/(4D)$ correspond to minimally and to conformally coupled
fields, respectively.

In addition to the field equation in the regions $r<a$ and $r>a$, the
matching conditions for the field should be specified at $r=a$. The field is
continuous on the separating surface: $\varphi |_{r=a-0}=\varphi |_{r=a+0}$.
In order to find the matching condition for the radial derivative of the
field, we note that the discontinuity of the functions $u^{\prime }(r)$ and $%
w^{\prime }(r)$ at $r=a$ leads to the delta function term%
\begin{equation}
2e^{-2v_{e}(a)}\sum_{j=i,e}\delta _{(j)}\left[ u_{j}^{\prime }(a-\delta
_{(j)}0)+(D-1)w_{j}^{\prime }(a-\delta _{(j)}0)\right] \delta (r-a)
\label{deltaR}
\end{equation}%
in the Ricci scalar and, hence, in the field equation (\ref{FieldEq}), if we
require its validity everywhere in the space. The expression (\ref{deltaR})
is given in terms of the trace of the surface energy-momentum tensor by
using the formula (\ref{Tracesurf}). As a result of the presence of the
delta function term in the field equation, the radial derivative of the
field has a discontinuity at $r=a$. The jump condition is obtained by
integrating the field equation through the point $r=a$. This gives%
\begin{equation}
\left( \partial _{r}\varphi \right) _{r=a+0}-\left( \partial _{r}\varphi
\right) _{r=a-0}=\frac{16\pi G\xi }{D-1}e^{v_{e}(a)}\tau \varphi |_{r=a}.
\label{DerJump}
\end{equation}%
For a minimally coupled field the radial derivative is continuous.

In what follows, we are interested in the VEVs of the field squared and of
the energy-momentum tensor induced in the region $r>a$ by the geometry in $%
r<a$. In the model under consideration all the information about the
properties of the vacuum is encoded in two-point functions. As such we will
use the positive frequency Wightman function defined as the VEV $%
W(x,x^{\prime })=\langle 0|\varphi (x)\varphi (x^{\prime })|0\rangle $,
where $|0\rangle $ stands for the vacuum state. In addition to describing
the local properties of the vacuum, this function also determines the
response of the Unruh-DeWitt type particle detectors \cite{Birr82}. For the
evaluation of the Wightman function we will use the direct summation over a
complete set of positive- and negative-energy mode functions $\left\{
\varphi _{\alpha }(x),\varphi _{\alpha }^{\ast }(x^{\prime })\right\} $,
obeying the field equation (\ref{FieldEq}) and the matching conditions
described above. Here, the set of quantum numbers $\alpha $ specifies the
solutions. Expanding the field operator over the complete set $\left\{
\varphi _{\alpha }(x),\varphi _{\alpha }^{\ast }(x^{\prime })\right\} $ and
using the standard commutation relations for the annihilation and creation
operators, the following mode-sum formula is readily obtained:
\begin{equation}
W(x,x^{\prime })=\sum_{\alpha }\varphi _{\alpha }(x)\varphi _{\alpha }^{\ast
}(x^{\prime }),  \label{Wsum}
\end{equation}%
where we assume summation over discrete quantum numbers and integration over
continuous ones.

In the problem under consideration, the mode functions can be presented in
the factorized form
\begin{equation}
\varphi _{\alpha }(x)=f_{l}(r)Y(m_{k};\vartheta ,\phi )e^{-i\omega t},\,
\label{modes}
\end{equation}%
where $l=0,1,2,\ldots $, $Y(m_{k};\vartheta ,\phi )$ is the hyperspherical
harmonic of degree $l$ \cite{Erde53}, $m_{k}=(m_{0}\equiv l,m_{1},\ldots
,m_{n})$, with $m_{1},m_{2},\ldots ,m_{n}$ being integers such that
\begin{equation}
0\leq m_{n-1}\leq m_{n-2}\leq \cdots \leq m_{1}\leq l,\quad -m_{n-1}\leq
m_{n}\leq m_{n-1}.  \label{mnumbvalues}
\end{equation}%
Presenting the radial function as%
\begin{equation}
f_{l}(r)=\left\{
\begin{array}{cc}
f_{(i)l}(r), & r<a, \\
f_{(e)l}(r) & r>a,%
\end{array}%
\right.   \label{fl}
\end{equation}%
the equations for the exterior and interior functions are obtained from (\ref%
{FieldEq})%
\begin{equation}
f_{(j)l}^{\prime \prime }(r)+\left[ u_{j}^{\prime }-v_{j}^{\prime
}+(D-1)w_{j}^{\prime }\right] f_{(j)l}^{\prime }(r)+e^{2v_{j}}\left[ \frac{%
\omega ^{2}}{e^{2u_{j}}}-m^{2}-\xi R_{(j)}-\frac{l(l+n)}{e^{2w_{j}}}\right]
f_{(j)l}(r)=0,  \label{fleq}
\end{equation}%
where the Ricci scalar is given by the expression (\ref{Richscin}). From the
matching conditions on the separating boundary, given above, for the radial
functions in the interior and exterior regions, we find $%
f_{(e)l}(a+0)=f_{(i)l}(a-0)$ and
\begin{equation}
f_{(e)l}^{\prime }(a+0)-f_{(i)l}^{\prime }(a-0)=\frac{16\pi G\xi }{D-1}%
e^{v_{e}(a)}\tau f_{(e)l}(a).  \label{flderjump}
\end{equation}%
Note that, introducing a new radial coordinate, the equation (\ref{fleq})
can be written in the Schr\"{o}dinger-like form%
\begin{equation}
\partial _{y}^{2}g_{(j)l}(y)+\left[ \omega ^{2}-U_{(j)l}(y)\right]
g_{(j)l}(y)=0,  \label{SchEq}
\end{equation}%
where%
\begin{equation}
g_{(j)l}(y)=e^{(D-1)w_{j}/2}f_{(j)l}(r),\;y=\int dr\,e^{v_{j}-u_{j}},
\label{gj}
\end{equation}%
and for the potential function we have%
\begin{equation}
U_{(j)l}(y)=e^{2u_{j}}\left[ m^{2}+\xi R_{(j)}+\frac{l(l+n)}{e^{2w_{j}}}%
\right] +\frac{D-1}{2}\left( w_{j}^{\prime \prime }+\frac{D-1}{2}%
w_{j}^{\prime 2}\right) .  \label{Uj}
\end{equation}

In what follows we assume that the interior geometry is regular. In terms of
the radial coordinate (\ref{rbar}), from the regularity of the Ricci scalar (%
\ref{Richscin}) at the center, $\bar{r}=0$, it follows that%
\begin{equation}
u_{i}(\bar{r}),v_{i}(\bar{r})\sim \bar{r}^{2},\;\bar{r}\rightarrow 0.
\label{RegCentr}
\end{equation}%
Let $f_{(i)\,l}^{(1)}(r,\lambda )$, with $\lambda ^{2}=\omega ^{2}-m^{2}$,
be the solution of the equation (\ref{fleq}) in the interior region which is
regular at the origin. It can be taken as a real function. In addition, by
taking into account that $\lambda $ enters in the equation in the form $%
\lambda ^{2}$, without loss of generality we can assume that $%
f_{(i)\,l}^{(1)}(r,-\lambda )=\mathrm{const}\cdot f_{(i)\,l}^{(1)}(r,\lambda
)$. From the regularity of the geometry at the center and from (\ref{fleq})
it follows that near the center the interior regular solution behaves as $%
f_{(i)\,l}^{(1)}(r,\lambda )\sim \bar{r}^{l}$.

Now, the radial parts of the mode functions are presented as%
\begin{equation}
f_{l}(r)=\left\{
\begin{array}{ll}
A_{(i)}f_{(i)\,l}^{(1)}(r,\lambda ), & \mathrm{for}\;r<a \\
A_{(e)1}f_{(e)\,l}^{(1)}(r,\lambda )+A_{(e)2}f_{(e)\,l}^{(2)}(r,\lambda ), &
\mathrm{for}\;r>a%
\end{array}%
,\right.  \label{radialsol}
\end{equation}%
where $f_{(e)\,l}^{(1)}(r,\lambda )$ and $f_{(e)\,l}^{(2)}(r,\lambda )$ are
the two linearly independent solutions of the radial equation in the
exterior region (equation (\ref{fleq}) with $j=e$). We assume that the
functions $f_{(e)\,l}^{(j)}(r,\lambda )$, $j=1,2$, are taken to be real. The
coefficients in (\ref{radialsol}) are determined by the continuity condition
for the radial functions and by the jump condition (\ref{flderjump}) for
their radial derivatives. From these conditions we get%
\begin{equation}
A_{(e)1}=A_{(i)}W_{l}^{(1)},\;A_{(e)2}=-A_{(i)}W_{l}^{(2)},  \label{Ae1}
\end{equation}%
with the notations%
\begin{eqnarray}
W_{l}^{(1)} &=&\frac{W_{l}^{(i2)}(a,\lambda )}{W_{l}^{(12)}(a)}-\frac{16\pi
G\xi }{D-1}e^{v_{e}(a)}\tau \frac{f_{(i)\,l}^{(1)}(a,\lambda )}{%
W_{l}^{(12)}(a)}f_{(e)\,l}^{(2)}(a,\lambda ),  \notag \\
W_{l}^{(2)} &=&\frac{W_{l}^{(i1)}(a,\lambda )}{W_{l}^{(12)}(a)}-\frac{16\pi
G\xi }{D-1}e^{v_{e}(a)}\tau \frac{f_{(i)\,l}^{(1)}(a,\lambda )}{%
W_{l}^{(12)}(a)}f_{(e)\,l}^{(1)}(a,\lambda ).  \label{Wl2}
\end{eqnarray}%
In (\ref{Wl2}) we have defined the functions%
\begin{eqnarray}
W_{l}^{(ij)}(r,\lambda ) &=&W\{f_{(i)\,l}^{(1)}(r,\lambda
),f_{(e)\,l}^{(j)}(r,\lambda )\},\;j=1,2  \notag \\
W_{l}^{(12)}(r) &=&W\{f_{(e)\,l}^{(1)}(r,\lambda
),f_{(e)\,l}^{(2)}(r,\lambda )\},  \label{Wl120}
\end{eqnarray}%
where $W\{f(r),g(r)\}=f(r)g^{\prime }(r)-f^{\prime }(r)g(r)$ is the
Wronskian. The Wronskian $W_{l}^{(12)}(r)$ can be found from the equation (%
\ref{fleq}) with $j=e$:%
\begin{equation}
W_{l}^{(12)}(r)=Ce^{-u_{e}(r)+v_{e}(r)-(D-1)w_{e}(r)},  \label{Wl12}
\end{equation}%
where the constant $C$ is determined by the choice of the functions $%
f_{(e)\,l}^{(1)}(r,\lambda )$ and $f_{(e)\,l}^{(2)}(r,\lambda )$. Here we
will assume that the exterior metric is asymptotically flat at large
distances from the boundary, $r\rightarrow \infty $. With this assumption,
we can see that for large $r$ the solution for the exterior equation is
given by $r^{-n}Z_{\nu _{l}}(\lambda r)$, where $Z_{\nu _{l}}(\lambda r)$ is
a cylinder function of the order%
\begin{equation}
\nu _{l}=l+n/2.  \label{nul}
\end{equation}%
If we take the functions $f_{(e)\,l}^{(1)}(r,\lambda )$ and $%
f_{(e)\,l}^{(2)}(r,\lambda )$ such that $f_{(e)\,l}^{(1)}(r,\lambda )\approx
r^{-n/2}J_{\nu _{l}}(\lambda r)$, $f_{(e)\,l}^{(2)}(r,\omega )\approx
r^{-n/2}Y_{\nu _{l}}(\lambda r)$, for $r\rightarrow \infty $, with $J_{\nu
}(x)$ and $Y_{\nu }(x)$ being the Bessel and the Neumann functions, then for
the constant in (\ref{Wl12}) we find $C=2/\pi $. In what follows we will
assume this choice of the normalization for the exterior mode functions. In
this way, as a complete set of quantum numbers specifying the mode functions
we can take the set $\alpha =(\lambda ,m_{k})$. Here we assume that $\lambda
$ is real. In addition, bound states can be present with purely imaginary $%
\lambda $. These states are discussed below.

The remaining coefficient $A_{(i)}$ is determined by the normalization
condition for the mode functions given by
\begin{equation}
\int d^{D}x\sqrt{|g|}g^{00}\varphi _{\alpha }(x)\overleftrightarrow{\partial
}_{t}\varphi _{\alpha ^{\prime }}^{\ast }(x)=i\delta \left( \lambda -\lambda
^{\prime }\right) \delta _{m_{k}m_{k}^{\prime }},  \label{Normaliz}
\end{equation}%
The integral over $r\leqslant a$ is finite and the divergence for $\lambda
=\lambda ^{\prime }$ comes from the upper limit of the integration over $r$.
As a consequence of this, we can replace the functions $f_{(e)\,l}^{(1)}(r,%
\omega )$ and $f_{(e)\,l}^{(2)}(r,\omega )$ by their asymptotics for $%
r\rightarrow \infty $. In this way, for the normalization coefficient one
finds%
\begin{equation}
A_{(i)}^{2}=\lambda \frac{(W_{l}^{(1)2}+W_{l}^{(2)2})^{-1}}{2N(m_{k})\omega }%
,  \label{Ai2}
\end{equation}%
with $W_{l}^{(1,2)}$ given by (\ref{Wl2}). Hence, for the radial
mode-functions we get%
\begin{equation}
f_{l}(r,\lambda )=A_{(i)}\left\{
\begin{array}{ll}
f_{(i)\,l}^{(1)}(r,\lambda ), & \mathrm{for}\;r<a \\
f_{(e)\,l}(r,\lambda ), & \mathrm{for}\;r>a%
\end{array}%
,\right.  \label{fl2}
\end{equation}%
where the notation
\begin{equation}
f_{(e)\,l}(r,\lambda )=W_{l}^{(1)}f_{(e)\,l}^{(1)}(r,\lambda
)-W_{l}^{(2)}f_{(e)\,l}^{(2)}(r,\lambda ),  \label{fel}
\end{equation}%
is introduced.

An equivalent form of the exterior mode functions is given by
\begin{equation}
f_{l}(r,\lambda )=A_{(e)}g_{l}(r,\lambda ),\;r>a,  \label{fle}
\end{equation}%
with the notation
\begin{equation}
g_{l}(r,\lambda )=\bar{f}_{(e)\,l}^{(2)}(a,\lambda
)f_{(e)\,l}^{(1)}(r,\lambda )-\bar{f}_{(e)\,l}^{(1)}(a,\lambda
)f_{(e)\,l}^{(2)}(r,\lambda ),  \label{gl}
\end{equation}%
and with the normalization coefficient
\begin{equation}
A_{(e)}^{2}=\lambda \frac{\lbrack \bar{f}_{(e)\,l}^{(1)2}(a,\lambda )+\bar{f}%
_{(e)\,l}^{(2)2}(a,\lambda )]^{-1}}{2N(m_{k})\omega }.  \label{Ae2}
\end{equation}%
Here and in what follows, for a given function $F(r,\lambda )$, we use the
notation%
\begin{equation}
\bar{F}(r,\lambda )=\partial _{r}F(r,\lambda )-\left[ \frac{%
f_{(i)\,l}^{(1)\prime }(a,\lambda )}{f_{(i)\,l}^{(1)}(a,\lambda )}+\frac{%
16\pi G\xi }{D-1}e^{v_{e}(a)}\tau \right] F(r,\lambda ),  \label{BarNot}
\end{equation}%
where $f_{(i)\,l}^{(1)\prime }(a,\lambda )=\partial
_{r}f_{(i)\,l}^{(1)}(r,\lambda )|_{r=a-0}$. Note that one has the relation%
\begin{equation}
A_{(e)}=A_{(i)}\frac{f_{(i)\,l}^{(1)}(a,\lambda )}{W_{l}^{(12)}(a)},
\label{Aei}
\end{equation}%
for the coefficients in the exterior and interior regions.

\subsection{Bound states}

In the previous subsection we have considered the modes with real $\lambda $%
. In addition to them, the modes with imaginary $\lambda $ can be present
which correspond to possible bound states. For these states the exterior
radial mode functions in the region $r\rightarrow \infty $ behave as $%
r^{-n/2}K_{\nu _{l}}(\eta r)$, where $\eta =|\lambda |$ and $K_{\nu }(x)$ is
the Macdonald function. In order to have a stable vacuum state we will
assume that $\eta <m$. For the radial functions corresponding to the bound
states one has
\begin{equation}
f_{bl}(r,\lambda )=\left\{
\begin{array}{ll}
A_{(ib)}f_{(ib)\,l}^{(1)}(r,\eta ), & \mathrm{for}\;r<a \\
A_{(eb)}f_{(eb)\,l}^{(2)}(r,\eta ), & \mathrm{for}\;r>a%
\end{array}%
,\right.   \label{fb}
\end{equation}%
where $f_{(eb)\,l}^{(2)}(r,\eta )\approx r^{-n/2}K_{\nu _{l}}(\eta r)$ for $%
r\rightarrow \infty $. The continuity of the mode functions at $r=a$ leads
to the relation%
\begin{equation}
A_{(ib)}f_{(ib)\,l}^{(1)}(a,\eta )=A_{(eb)}f_{(eb)\,l}^{(2)}(a,\eta ).
\label{Contb1}
\end{equation}%
From the jump condition for the radial derivative we see that the allowed
values of $\eta $ for bound states are solutions of the equation%
\begin{equation}
\widehat{f}_{(eb)\,l}^{(2)}(a,\eta )=0,  \label{BoundSt}
\end{equation}%
where for a function $F(r,\eta )$ we define
\begin{equation}
\widehat{F}(r,\eta )=\partial _{r}F(r,\eta )-\left[ \frac{%
f_{(ib)\,l}^{(1)\prime }(a,\eta )}{f_{(ib)\,l}^{(1)}(a,\eta )}+\frac{16\pi
G\xi }{D-1}e^{v_{e}(a)}\tau \right] F(r,\eta ).  \label{Fhat}
\end{equation}%
The possible solutions of the equation (\ref{BoundSt}) will be denoted by $%
\eta =\eta _{s}$, $s=1,2,\ldots $.

The remaining coefficient in the mode functions (\ref{fb}) is determined
from the normalization condition for the bound states%
\begin{eqnarray}
A_{(eb)}^{-2} &=&2\omega N(m_{k})\bigg[\int_{a}^{\infty
}dr\,e^{-u_{e}+v_{e}+(D-1)w_{e}}f_{(eb)\,l}^{(2)2}(r,\eta )  \notag \\
&&+\frac{f_{(eb)\,l}^{(2)2}(a,\eta )}{f_{(i)\,l}^{2}(a,\eta )}%
\int_{r_{c}}^{a}dr\,e^{-u_{i}+v_{i}+(D-1)w_{i}}f_{(ib)\,l}^{(1)2}(r,\eta )%
\bigg],  \label{NormBS}
\end{eqnarray}%
with $\eta =\eta _{s}$. In order to evaluate the integrals in this formula
we note that for a solution $f_{(j)\omega l}(r)$ to the radial equation (\ref%
{fleq}) the following formula can be proved:%
\begin{eqnarray}
&&\int dr\,e^{-u_{j}+v_{j}+(D-1)w_{j}}f_{(j)\omega l}(r)f_{(j)\omega
_{1}l}(r)=\frac{e^{u_{j}-v_{j}+(D-1)w_{j}}}{\omega _{1}^{2}-\omega ^{2}}
\notag \\
&&\qquad \times \left[ f_{(j)\omega l}^{\prime }(r)f_{(j)\omega
_{1}l}(r)-f_{(j)\omega l}(r)f_{(j)\omega _{1}l}^{\prime }(r)\right] .
\label{IntegModes}
\end{eqnarray}%
In particular, in the limit $\omega _{1}\rightarrow \omega $, from here one
can obtain%
\begin{equation}
\int dr\,e^{-u_{j}+v_{j}+(D-1)w_{j}}f_{(j)\omega l}^{2}(r)=\frac{%
e^{u_{j}-v_{j}+(D-1)w_{j}}}{2\omega }\left[ f_{(j)\omega l}^{\prime
}(r)\partial _{\omega }f_{(j)\omega l}(r)-f_{(j)\omega l}(r)\partial
_{\omega }f_{(j)\omega l}^{\prime }(r)\right] .  \label{appintform1}
\end{equation}

Applying to the integrals in Eq. (\ref{NormBS}) the formula (\ref%
{appintform1}) with $\omega =\sqrt{m^{2}-\eta ^{2}}$ and using the
continuity of the radial eigenfunctions at $r=a$, for the normalization
coefficient one finds%
\begin{equation}
A_{(eb)}^{-2}=N(m_{k})e^{u(a)-v(a)+(D-1)w(a)}f_{(eb)\,l}^{(2)}(a,\eta
)\partial _{\omega }\widehat{f}_{(eb)\,l}^{(2)}(a,\eta ).  \label{Aeb}
\end{equation}%
The coefficient $A_{(i)}$ is found from (\ref{Contb1}).

An equivalent expression for the normalization coefficient is obtained by
using the Wronskian relation
\begin{equation}
f_{(eb)\,l}^{(2)}(a,\eta )f_{(eb)\,l}^{(1)\prime }(a,\eta
)-f_{(eb)\,l}^{(1)}(a,\eta )f_{(eb)\,l}^{(2)\prime }(a,\eta
)=e^{-u_{e}(a)+v_{e}(a)-(D-1)w_{e}(a)},  \label{WronskBS}
\end{equation}%
for two linearly independent solutions of the radial equation in the
exterior region. Here, the function $f_{(eb)\,l}^{(1)}(r,\eta )$ is
normalized by the relation $f_{(eb)\,l}^{(1)}(r,\eta )\approx r^{-n/2}I_{\nu
_{l}}(\eta r)$ for $r\rightarrow \infty $, with $I_{\nu }(x)$ being the
modified Bessel function. From (\ref{WronskBS}) we get%
\begin{equation}
f_{(eb)\,l}^{(2)}(a,\eta )\widehat{f}_{(eb)\,l}^{(1)}(a,\eta
)-f_{(eb)\,l}^{(1)}(a,\eta )\widehat{f}_{(eb)\,l}^{(2)}(a,\eta
)=e^{-u_{e}(a)+v_{e}(a)-(D-1)w_{e}(a)}.  \label{WronskBS2}
\end{equation}%
By taking into account that for the bound states one has the equation (\ref%
{BoundSt}), this gives%
\begin{equation*}
f_{(eb)\,l}^{(2)}(a,\eta _{s})=\frac{e^{-u_{e}(a)+v_{e}(a)-(D-1)w_{e}(a)}}{%
\widehat{f}_{(eb)\,l}^{(1)}(a,\eta _{s})}.
\end{equation*}%
Hence, the normalization constant for the exterior modes is written in the
form
\begin{equation}
A_{(eb)}^{2}=-\frac{\eta \widehat{f}_{(eb)\,l}^{(1)}(a,\eta )}{%
N(m_{k})\omega \partial _{\eta }\widehat{f}_{(eb)\,l}^{(2)}(a,\eta )},
\label{Ae2b}
\end{equation}%
with $\eta =\eta _{s}$.

\section{Wightman function}

\label{sec:WF}

Having a complete set of modes we can proceed to the evaluation of the
Wightman function by using the mode sum formula (\ref{Wsum}). First we
consider the case with no bound states. Substituting the functions (\ref%
{modes}) in (\ref{Wsum}), the summation over $m_{k}$ is done by using the
addition formula for the hyperspherical harmonics \cite{Erde53}
\begin{equation}
\sum_{m_{k}}\frac{Y(m_{k};\vartheta ,\phi )}{N(m_{k})}Y^{\ast
}(m_{k};\vartheta ^{\prime },\phi ^{\prime })=\frac{2l+n}{nS_{D}}%
C_{l}^{n/2}(\cos \theta ),  \label{YSum}
\end{equation}%
where $\theta $ is the angle between the directions determined by the angles
$(\vartheta ,\phi )$ and $(\vartheta ^{\prime },\phi ^{\prime })$. In (\ref%
{YSum}), $S_{D}=2\pi ^{D/2}/\Gamma (D/2)$ is the surface area of the unit
sphere in $D$-dimensional space and $C_{l}^{p}(x)$ is the Gegenbauer
polynomial of degree $l$ and order $p$. With the modes (\ref{fle}) and the
normalization coefficient (\ref{Ae2}), the expression for the Wightman
function in the exterior region reads:%
\begin{equation}
W(x,x^{\prime })=\sum_{l=0}^{\infty }\frac{l+n/2}{nS_{D}}C_{l}^{n/2}(\cos
\theta )\int_{0}^{\infty }d\lambda \,\frac{\lambda }{\omega }\frac{%
g_{l}(r,\lambda )g_{l}(r^{\prime },\lambda )e^{-i\omega \Delta t}}{\bar{f}%
_{(e)\,l}^{(1)2}(a,\lambda )+\bar{f}_{(e)\,l}^{(2)2}(a,\lambda )}.
\label{WFext}
\end{equation}%
where $\Delta t=t-t^{\prime }$ and the function $g_{l}(r,\lambda )$ is
defined by (\ref{gl}).

In order to separate from the Wightman function the contribution induced by
the interior geometry, firstly we introduce the functions
\begin{equation}
f_{(e)\,l}^{(\pm )}(r,\lambda )=f_{(e)\,l}^{(1)}(r,\lambda )\pm
if_{(e)\,l}^{(2)}(r,\lambda ).  \label{feplm}
\end{equation}%
Note that, as the functions $f_{(e)\,l}^{(1)}(r,\lambda )$ and $%
f_{(e)\,l}^{(2)}(r,\lambda )$ are real, one has $f_{(e)\,l}^{(-)}(r,\lambda
)=f_{(e)\,l}^{(+)\ast }(r,\lambda )$. For these new functions, at large
distances, $r\gg a$, one has the asymptotics%
\begin{equation}
f_{(e)\,l}^{(+)}(r,\lambda )\approx r^{-n/2}H_{\nu _{l}}^{(1)}(\lambda
r),\;f_{(e)\,l}^{(-)}(r,\lambda )\approx r^{-n/2}H_{\nu _{l}}^{(2)}(\lambda
r),  \label{feplmas}
\end{equation}%
with $H_{\nu _{l}}^{(1,2)}(x)$ being the Hankel functions. Now, it can be
seen that the following identity takes place:%
\begin{equation}
\frac{g_{l}(r,\lambda )g_{l}(r^{\prime },\lambda )}{\bar{f}%
_{(e)\,l}^{(1)2}(a,\lambda )+\bar{f}_{(e)\,l}^{(2)2}(a,\lambda )}%
=f_{(e)\,l}^{(1)}(r,\lambda )f_{(e)\,l}^{(1)}(r^{\prime },\lambda )-\frac{1}{%
2}\sum_{j=+,-}\frac{\bar{f}_{(e)\,l}^{(1)}(a,\lambda )}{\bar{f}%
_{(e)\,l}^{(j)}(a,\lambda )}f_{(e)\,l}^{(j)}(r,\lambda
)f_{(e)\,l}^{(j)}(r^{\prime },\lambda ).  \label{Ident}
\end{equation}

By using the relation (\ref{Ident}), the Wightman function from (\ref{WFext}%
) can be written in the decomposed form:%
\begin{equation}
W(x,x^{\prime })=W_{0}(x,x^{\prime })+W_{\mathrm{c}}(x,x^{\prime }),
\label{Wdec}
\end{equation}%
with the functions%
\begin{equation}
W_{0}(x,x^{\prime })=\sum_{l=0}^{\infty }\frac{l+n/2}{nS_{D}}%
C_{l}^{n/2}(\cos \theta )\int_{0}^{\infty }d\lambda \,\frac{\lambda }{\omega
}f_{(e)\,l}^{(1)}(r,\lambda )f_{(e)\,l}^{(1)}(r^{\prime },\lambda
)e^{-i\omega \Delta t},  \label{W0}
\end{equation}%
and%
\begin{eqnarray}
W_{\mathrm{c}}(x,x^{\prime }) &=&-\sum_{l=0}^{\infty }\frac{l+n/2}{2nS_{D}}%
C_{l}^{n/2}(\cos \theta )\sum_{j=+,-}\int_{0}^{\infty }d\lambda \,\frac{%
\lambda }{\omega }  \notag \\
&&\times \frac{\bar{f}_{(e)\,l}^{(1)}(a,\lambda )}{\bar{f}%
_{(e)\,l}^{(j)}(a,\lambda )}f_{(e)\,l}^{(j)}(r,\lambda
)f_{(e)\,l}^{(j)}(r^{\prime },\lambda )e^{-i\omega \Delta t}.  \label{Wc}
\end{eqnarray}%
The function $W_{0}(x,x^{\prime })$ is the Wightman function in the case of
the background when the geometry is described by the line element (\ref%
{metricoutside}) for all values of the radial coordinate $r$. As a radial
function in the corresponding modes the function $f_{(e)\,l}^{(1)}(r,\lambda
)$ is taken. Recall that we have $f_{(e)\,l}^{(1)}(r,\lambda )\approx
r^{-n/2}J_{\nu _{l}}(\lambda r)$ for $r\rightarrow \infty $ and, hence, for
these modes the vacuum state at asymptotic infinity coincides with the
Minkowskian vacuum. Thus, the function $W_{\mathrm{c}}(x,x^{\prime })$ can
be interpreted as the contribution to the Wightman function induced by the
geometry in the region $r<a$ with the line element (\ref{metricinside}).

If bound states are present, the contribution of the corresponding modes to
the Wightman function should be added to (\ref{Wdec}). For this
contribution, by using the mode functions (\ref{fb}) with the normalization
coefficient (\ref{Ae2b}), in the exterior region we get%
\begin{equation}
W_{\mathrm{bs}}(x,x^{\prime })=-\sum_{l=0}^{\infty }\frac{2l+n}{nS_{D}}%
C_{l}^{n/2}(\cos \theta )\sum_{\eta =\eta _{s}}\frac{\eta \widehat{f}%
_{(eb)\,l}^{(1)}(a,\eta )}{\omega \partial _{\eta }\widehat{f}%
_{(eb)\,l}^{(2)}(a,\eta )}f_{(eb)\,l}^{(2)}(r,\eta
)f_{(eb)\,l}^{(2)}(r^{\prime },\eta )e^{-i\omega \Delta t},  \label{Wbs}
\end{equation}%
where $\omega =\sqrt{m^{2}-\eta ^{2}}$ and $\eta =\eta _{s}$ are solutions
of the equation (\ref{BoundSt}).

The part $W_{\mathrm{c}}(x,x^{\prime })$ of the Wightman function, induced
by the interior geometry, can be further transformed by taking into account
that, for large values of $\lambda $, for the functions $f_{(e)\,l}^{(j)}(r,%
\lambda )$ in (\ref{Wc}) one has $f_{(e)\,l}^{(j)}(r,\lambda )\sim
e^{ji\lambda r}$. By using this property and under the condition $|\Delta
t|<(r+r^{\prime }-2a)$, assuming that the function $f_{(e)\,l}^{(+)}(r,%
\lambda )$ ($f_{(e)\,l}^{(-)}(r,\lambda )$) has no zeros for $0<\arg \lambda
<\pi /2$ ($-\pi /2<\arg \lambda <0$), in (\ref{Wc}) we can rotate the
integration contour in the complex plane $\lambda $ by the angle $\pi /2$ ($%
-\pi /2$) for the term with $j=+$ ($j=-$). In the presence of bound states,
the integrands have simple poles at $\lambda =\eta _{s}e^{j\pi i/2}$,
corresponding to the zeros of the function $\bar{f}_{(e)\,l}^{(j)}(a,\lambda
)$ on the imaginary axis. These poles have to be circled on the right along
contours with small radii. In the integrals over the imaginary axis ($%
\lambda =\eta e^{\pm i\pi /2}$), the integrands are expressed in terms of
the functions $f_{(e)\,l}^{(1)}(a,\eta e^{\pm \pi i/2})$ and $%
f_{(e)\,l}^{(j)}(r,\eta e^{ji\pi /2})$, $j=+,-$. By comparing the
asypmtotics of the functions for $r\rightarrow \infty $, we can see that the
functions $f_{(e)\,l}^{(j)}(r,\eta e^{j\pi i/2})$ are reduced to the
function $f_{(eb)\,l}^{(2)}(r,\eta )$, up to a coefficient and the function $%
f_{(e)\,l}^{(1)}(a,\eta e^{\pi i/2})$ is reduced to the function $%
f_{(eb)\,l}^{(1)}(a,\eta )$. By taking into account the normalization of the
functions for large $r$ and the relations between the Bessel functions and
modified Bessel functions, we conclude that%
\begin{eqnarray}
f_{(e)\,l}^{(j)}(r,\eta e^{j\pi i/2}) &=&-j\frac{2i}{\pi }e^{-j\nu _{l}\pi
i/2}f_{(eb)\,l}^{(2)}(r,\eta ),  \notag \\
f_{(e)\,l}^{(1)}(a,\eta e^{j\pi i/2}) &=&e^{j\nu _{l}\pi
i/2}f_{(eb)\,l}^{(1)}(a,\eta ).  \label{FuncRel}
\end{eqnarray}%
By using these relations, one can see that the integrals over the regions $%
(0,im)$ and $(0,-im)$ cancel out, whereas the integrals over small
semicircles around the poles $\eta _{s}e^{j\pi i/2}$ combine in the residue
at the point $\eta _{s}e^{\pi i/2}$. An interesting fact is that the
contribution of this residue to the part of the Wightman function (\ref{Wc})
exactly cancels the corresponding contribution coming from the bound state
(see (\ref{Wbs})). Finally, we get the following representation:%
\begin{eqnarray}
W_{\mathrm{c}}(x,x^{\prime }) &=&-\sum_{l=0}^{\infty }\frac{2l+n}{\pi nS_{D}}%
C_{l}^{n/2}(\cos \theta )\int_{m}^{\infty }d\eta \,\eta \frac{\widehat{f}%
_{(eb)\,l}^{(1)}(a,\eta )}{\widehat{f}_{(eb)\,l}^{(2)}(a,\eta )}  \notag \\
&&\times \frac{\cosh (\Delta t\sqrt{\eta ^{2}-m^{2}})}{\sqrt{\eta ^{2}-m^{2}}%
}f_{(eb)\,l}^{(2)}(r,\eta )f_{(eb)\,l}^{(2)}(r^{\prime },\eta ).
\label{Wdec2}
\end{eqnarray}%
Recall that, in deriving this formula we have assumed that $|\Delta
t|<(r+r^{\prime }-2a)$. In particular, this condition is obeyed in the
coincidence limit. An important advantage of the representation (\ref{Wdec2}%
), compared with (\ref{Wc}), is that in the upper limit of the integration
the integrand decays exponentially instead of the strongly oscillatory
behavior in (\ref{Wc}).

With the known Wightman function, we can evaluate the VEVs of the field
squared and the energy-momentum tensor by using the formulae below:%
\begin{eqnarray}
\langle 0|\varphi ^{2}|0\rangle  &=&\lim_{x^{\prime }\rightarrow
x}W(x,x^{\prime }),  \notag \\
\langle 0|T_{ik}|0\rangle  &=&\lim_{x^{\prime }\rightarrow x}\partial
_{i}\partial _{k}^{\prime }W(x,x^{\prime })+\left[ \left( \xi -1/4\right)
g_{ik}\nabla _{l}\nabla ^{l}-\xi \nabla _{i}\nabla _{k}-\xi R_{ik}\right]
\langle 0|\varphi ^{2}|0\rangle ,  \label{Tik}
\end{eqnarray}%
where $R_{ik}$ is the Ricci tensor for the background spacetime. The
expression for the energy--momentum tensor in (\ref{Tik}) differs from the
standard one, given, for example, in \cite{Birr82}, by the term which
vanishes on the mass shell (see \cite{Saha04}). Similarly to the Wightman
function, the VEVs are decomposed into two parts%
\begin{eqnarray}
\langle 0|\varphi ^{2}|0\rangle  &=&\langle \varphi ^{2}\rangle _{0}+\langle
\varphi ^{2}\rangle _{\mathrm{c}},  \notag \\
\langle 0|T_{ik}|0\rangle  &=&\langle T_{ik}\rangle _{0}+\langle
T_{ik}\rangle _{\mathrm{c}},  \label{Tikdec}
\end{eqnarray}%
where the parts $\langle \varphi ^{2}\rangle _{0}$ and $\langle
T_{ik}\rangle _{0}$ are obtained from the Wightman function $%
W_{0}(x,x^{\prime })$. The contributions $\langle \varphi ^{2}\rangle _{%
\mathrm{c}}$ and $\langle T_{ik}\rangle _{\mathrm{c}}$ are induced by the
geometry in the region $r<a$ and are given by formulae similar to (\ref{Tik}%
) with $W(x,x^{\prime })$ replaced by $W_{\mathrm{c}}(x,x^{\prime })$.

Of course, the coincidence limits in (\ref{Tik}) are divergent and a
renormalization procedure is necessary. An important point to be mentioned
here is that, for points $r>a$ the local geometry is not changed by the
interior region and, as a consequence, the divergences are contained in the
parts $\langle \varphi ^{2}\rangle _{0}$ and $\langle T_{ik}\rangle _{0}$
and the parts $\langle \varphi ^{2}\rangle _{\mathrm{c}}$ and $\langle
T_{ik}\rangle _{\mathrm{c}}$ are finite. Hence, providing an explicit
decomposition of the Wightman function in the form (\ref{Wdec2}), we have
reduced the renormalization procedure for the VEVs to the one in the case of
the background where the geometry is described by the line element (\ref%
{metricoutside}) for all values of the radial coordinate $r$.

In particular, by taking into account that%
\begin{equation}
C_{l}^{n/2}(1)=\frac{\Gamma (l+n)}{\Gamma (n)\Gamma (l+1)},  \label{Cl1}
\end{equation}%
for the contribution in the VEV of the field squared induced by the interior
geometry we get the expression%
\begin{equation}
\langle \varphi ^{2}\rangle _{\mathrm{c}}=-\frac{\Gamma (D/2)}{2\pi ^{D/2+1}}%
\sum_{l=0}^{\infty }D_{l}\int_{m}^{\infty }d\eta \,\frac{\widehat{f}%
_{(eb)\,l}^{(1)}(a,\eta )}{\widehat{f}_{(eb)\,l}^{(2)}(a,\eta )}\frac{\eta
f_{(eb)\,l}^{(2)2}(r,\eta )}{\sqrt{\eta ^{2}-m^{2}}},  \label{phi2Gen}
\end{equation}%
where%
\begin{equation}
D_{l}=\frac{(2l+n)\Gamma (l+n)}{\Gamma (D-1)\Gamma (l+1)},  \label{Dl1}
\end{equation}%
is the degeneracy of the angular mode with given $l$. The corresponding VEV
of the energy-momentum tensor is obtained from (\ref{Tik}). The VEV (\ref%
{phi2Gen}), in general, diverges on the boundary, $r=a$. The leading term of
the corresponding asymptotic expansion over the distance from the boundary
depends on the specific interior and exterior geometries and examples will
be done below.

We have considered the Wightman function in the exterior region. The mode
sum for the corresponding function in the interior region is obtained by
using the interior modes from (\ref{radialsol}) with the normalization
coefficient (\ref{Ai2}). Subtracting from the mode sum the Wightman function
for the geometry described by the line element (\ref{metricinside}) for all
values of the radial coordinate, we can separate the part induced by the
exterior geometry. In what follows we will be concerned with the VEVs in the
exterior region.

\section{Minkowski spacetime as an exterior geometry}

\label{sec:ExtMink}

\subsection{Wightman function}

As an application of general results given in previous sections, here we
assume that the exterior geometry is described by the Minkowski spacetime.
The corresponding line element has the form
\begin{equation}
ds_{e}^{2}=dt^{2}-dr^{2}-r^{2}d\Omega _{D-1}^{2},  \label{ds2eM}
\end{equation}%
with the functions appearing in (\ref{metricoutside}):
\begin{equation}
u_{e}(r)=v_{e}(r)=0,\;e^{w_{e}(r)}=r.  \label{ueMink}
\end{equation}%
In this case, in the exterior region we have the radial functions%
\begin{eqnarray}
f_{(e)\,l}^{(1)}(r,\lambda ) &=&r^{-n/2}J_{\nu _{l}}(\lambda
r),\;f_{(e)\,l}^{(2)}(r,\lambda )=r^{-n/2}Y_{\nu _{l}}(\lambda r),  \notag \\
f_{(e)\,l}^{(+)}(r,\lambda ) &=&r^{-n/2}H_{\nu _{l}}^{(1)}(\lambda
r),\;f_{(e)\,l}^{(-)}(r,\lambda )=r^{-n/2}H_{\nu _{l}}^{(2)}(\lambda r).
\label{feMink}
\end{eqnarray}%
For the corresponding functions on the imaginary axis we get%
\begin{equation}
f_{(eb)\,l}^{(1)}(r,\eta )=r^{-n/2}I_{\nu _{l}}(\eta
r),\;f_{(eb)\,l}^{(2)}(r,\eta )=r^{-n/2}K_{\nu _{l}}(\eta r).
\label{feMink2}
\end{equation}

In the special case under consideration, $W_{0}(x,x^{\prime })$ is the
Wightman function in the Minkowski spacetime. For the contribution induced
by the interior geometry we have the expression
\begin{eqnarray}
W_{\mathrm{c}}(x,x^{\prime }) &=&-\sum_{l=0}^{\infty }\frac{\left(
2l+n\right) C_{l}^{n/2}(\cos \theta )}{\pi nS_{D}\left( rr^{\prime }\right)
^{n/2}}\int_{m}^{\infty }d\eta \,\eta \frac{\tilde{I}_{\nu _{l}}(\eta a)}{%
\tilde{K}_{\nu _{l}}(\eta a)}  \notag \\
&&\times \frac{\cosh (\Delta t\sqrt{\eta ^{2}-m^{2}})}{\sqrt{\eta ^{2}-m^{2}}%
}K_{\nu _{l}}(\eta r)K_{\nu _{l}}(\eta r^{\prime }).  \label{WMExt2}
\end{eqnarray}%
In this formula, for a given function $F(z)$ we have defined
\begin{equation}
\tilde{F}(z)=zF^{\prime }(z)-\left[ ay_{l}(a,\eta )+\frac{16\pi G\xi }{D-1}%
\tau a+\frac{D}{2}-1\right] F(z),  \label{Ftilde1}
\end{equation}%
with the notation
\begin{equation}
y_{l}(r,\eta )=\frac{\partial _{r}f_{(i)l}^{(1)}(r,i\eta )}{%
f_{(i)l}^{(1)}(r,i\eta )},  \label{yi}
\end{equation}%
and with
\begin{equation}
\frac{8\pi G}{D-1}\tau =u_{i}^{\prime }(a)+(D-1)[w_{i}^{\prime }(a)-1/a].
\label{tauM}
\end{equation}%
Note that, for the exterior Minkowskian spacetime, the equation (\ref%
{BoundSt}), defining the bound states, is written in the form%
\begin{equation}
\tilde{K}_{\nu _{l}}(\eta a)=0,\;\eta <m.  \label{BoundStM}
\end{equation}%
The existence of the solutions for this equation with $\eta >m$ leads to the
instability of the exterior Minkowskian vacuum. An example of this type will
be discussed below in Section \ref{sec:dS}.

The expression (\ref{WMExt2}) differs from the corresponding formula for the
Wightman function outside a spherical boundary in Minkowski spacetime with
Robin boundary condition $\left( \beta _{\mathrm{R}}+\partial _{r}\right)
\varphi =0$ at $r=a$ (see \cite{Saha01}), by the replacement of the Robin
coefficient%
\begin{equation}
\beta _{\mathrm{R}}\rightarrow -y_{l}(a,\eta )-\frac{16\pi G\xi }{D-1}\tau .
\label{Rob}
\end{equation}%
In the problem under consideration, the effective Robin coefficient depends
on both $\eta $ and $l$. As it will be shown below, this leads to the
weakening of divergences in the local VEVs on the boundary.

\subsection{VEV of the field squared}

The renormalization of the VEVs in the exterior region is reduced to the
subtraction of the corresponding VEVs in Minkowski spacetime. In this case
the renormalized VEVs of the field squared and the energy-momentum tensor
coincide with the parts $\left\langle \varphi ^{2}\right\rangle _{\mathrm{c}}
$ and $\langle T_{ik}\rangle _{\mathrm{c}}$ induced by the interior
geometry. For the renormalized VEV of the field squared we get%
\begin{equation}
\left\langle \varphi ^{2}\right\rangle _{\mathrm{c}}=-\frac{\Gamma (D/2)}{%
2\pi ^{D/2+1}r^{D-2}}\sum_{l=0}^{\infty }D_{l}\int_{m}^{\infty }d\eta \,\eta
\frac{\tilde{I}_{\nu _{l}}(a\eta )}{\tilde{K}_{\nu _{l}}(a\eta )}\frac{%
K_{\nu _{l}}^{2}(r\eta )}{\sqrt{\eta ^{2}-m^{2}}}.  \label{phi2}
\end{equation}%
Let us discuss the behavior of this VEV in the asymptotic regions of the
parameters.

At large distances from the boundary and for a massive filed, assuming that $%
mr\gg 1$ for a fixed $ma$, the dominant contribution to the integral in (\ref%
{phi2}) comes from the region near the lower limit of the integration. By
using the asymptotic formula for the Macdonald function for large values of
the argument, to the leading order we get%
\begin{equation}
\left\langle \varphi ^{2}\right\rangle _{\mathrm{c}}\approx -\frac{\Gamma
(D/2)e^{-2rm}}{8\pi ^{(D+1)/2}\sqrt{mr}r^{D-1}}\sum_{l=0}^{\infty }D_{l}%
\frac{\tilde{I}_{\nu _{l}}(am)}{\tilde{K}_{\nu _{l}}(am)}.  \label{phi2Large}
\end{equation}%
Hence, at distances from the boundary larger than the Compton wavelength,
the VEV is exponentially suppressed. For a massless field and for $r\gg a$,
we introduce in (\ref{phi2}) a new integration variable $y=r\eta $ and
expand the functions $\tilde{I}_{\nu _{l}}(ya/r)$ and $\tilde{K}_{\nu
_{l}}(ya/r)$. The contribution of the leading term for a given $l$ behaves
as $(a/r)^{2l+2D-3}$ and the integral is evaluated by using the formula%
\begin{equation}
\mathcal{I}(\nu )\equiv \int_{0}^{\infty }dy\,y^{2\nu }K_{\nu }^{2}(y)=\frac{%
\pi \Gamma \left( 2\nu +1/2\right) \Gamma \left( \nu +1/2\right) }{4\Gamma
(\nu +1)}.  \label{IntK2}
\end{equation}%
The dominant contribution comes from the term with the lowest orbital
momentum $l=0$ with the leading term%
\begin{equation}
\left\langle \varphi ^{2}\right\rangle _{\mathrm{c}}\approx \frac{%
D/2-1-\beta _{0}}{D/2-1+\beta _{0}}\frac{(D-2)\Gamma (D-3/2)\Gamma ((D-1)/2)%
}{2(4\pi )^{D/2}\Gamma ^{2}(D/2)a^{D-1}}(a/r)^{2D-3},  \label{phi2Largem0}
\end{equation}%
where%
\begin{equation}
\beta _{0}=ay_{0}(a,0)+\frac{16\pi G\xi }{D-1}\tau a+\frac{D}{2}-1.
\label{bet0}
\end{equation}%
If $\beta _{0}=D/2-1$ or $\beta _{0}=1-D/2$, the next-to-leading order terms
should be kept in the expansions of the functions $\tilde{I}_{\nu _{l}}(ya/r)
$ and $\tilde{K}_{\nu _{l}}(ya/r)$, respectively. Hence, for a massless
field the decay of the VEV at large distances follows a power-law.

The VEV of the field squared (\ref{phi2}) diverges on the boundary, $r=a$.
The surface divergences in the VEVs of local physical observables are
well-known in the theory of the Casimir effect and were investigated for
various types of boundary geometries. In the problem at hand, the appearance
of divergences is related to the idealized model of the zero thickness
transition range between the interior and exterior geometries. In order to
find the leading term in the asymptotic expansion over the distance from the
boundary, we note that for points near the boundary the dominant
contribution to the series in (\ref{phi2}) comes from large values of $l$.
For these $l$, introducing a new integration variable $x=a\eta /\nu _{l}$,
we use the uniform asymptotic expansions for the modified Bessel functions
for large values of the order (see, for instance, \cite{Abra72}). For the
further evaluation we need also the uniform asymptotic expansion of the
function $y_{l}(r,\eta )$. From the equation (\ref{fleq}) for the interior
radial mode function the following equation for the function (\ref{yi}) is
obtained:%
\begin{eqnarray}
&&y_{l}^{\prime }(r,\eta )+y_{l}^{2}(r,\eta )+\left[ u_{i}^{\prime
}-v_{i}^{\prime }+(D-1)w_{i}^{\prime }\right] y_{l}(r,\eta )  \notag \\
&&\qquad -e^{2v_{i}}\left[ \frac{\eta ^{2}-m^{2}}{e^{2u_{i}}}+m^{2}+\xi
R_{(i)}+\frac{l(l+n)}{e^{2w_{i}}}\right] =0.  \label{yiEq}
\end{eqnarray}%
From here it follows that for the leading term in the asymptotic expansion
of the function $y_{l}(r,\nu _{l}x)$ for large values of $l$ one has%
\begin{equation}
y_{l}(r,\nu _{l}x)\approx \pm \nu _{l}e^{v_{i}}\sqrt{%
e^{-2u_{i}}x^{2}+e^{-2w_{i}}}.  \label{ylr}
\end{equation}%
For the function $f_{(i)l}(r,i\eta )$ in (\ref{yi}), regular at the center,
the upper sign should be taken. By taking into account that $%
u_{i}(a)=v_{i}(a)=0$ and $e^{2w_{i}(a)}=a^{2}$, the asymptotic expansion at $%
r=a$ can be written as%
\begin{equation}
y_{l}(a,\nu _{l}x)\approx \frac{\nu _{l}}{a}\sqrt{a^{2}x^{2}+1}\left[ 1+%
\frac{B(ax)}{\nu _{l}}+\cdots \right] ,  \label{yia}
\end{equation}%
where the function $B(ax)$ depends on the interior geometry.

By making use of the uniform asymptotic expansions for the modified Bessel
functions, with the combination of (\ref{yia}), we can see that the leading
order contribution to the function $\tilde{I}_{\nu _{l}}(\nu _{l}x)$ coming
form the first term in (\ref{yia}) is cancelled by the leading term in the
asymptotic expansion of the function $zI_{\nu _{l}}^{\prime }(z)$ with $%
z=\nu _{l}x$. As a result, for the ratio appearing in (\ref{phi2}), in the
leading order, we get%
\begin{equation}
\frac{\tilde{I}_{\nu _{l}}(\nu _{l}x)}{\tilde{K}_{\nu _{l}}(\nu _{l}x)}%
\approx \frac{C(x)}{2\pi l}e^{2l\eta (x)},  \label{IKtilde}
\end{equation}%
with the function%
\begin{equation}
C(x)=B(x)+\left( \frac{16\pi G\xi }{D-1}\tau a+\frac{D}{2}-1\right) \frac{1}{%
\sqrt{1+x^{2}}}+\frac{x^{2}/2}{\left( 1+x^{2}\right) ^{3/2}},  \label{Cx}
\end{equation}%
and with the standard notation (see \cite{Abra72})%
\begin{equation}
\eta (x)=\sqrt{1+x^{2}}+\ln \frac{x}{1+\sqrt{1+x^{2}}}.  \label{etax}
\end{equation}%
The function $B(x)$ for special cases of the interior dS and AdS spaces will
be given below.

Substituting (\ref{IKtilde}) and the uniform asymptotic expansion for the
function $K_{\nu _{l}}^{2}(\nu _{l}xr/a)$ into (\ref{phi2}), with a new
integration variable $x=a\eta /\nu _{l}$, in the leading order we use the
relations $D_{l}\approx 2l^{D-2}/\Gamma (D-1)$ and $\eta (xr/a)-\eta
(x)\approx \sqrt{1+x^{2}}(r/a-1)$. In the same order, by taking into account
that $\sum_{l=0}^{\infty }l^{p-1}e^{-\alpha l}\approx \Gamma (p)/\alpha ^{p}$
for $\alpha \rightarrow 0$, for the leading term in the asymptotic expansion
of the VEV for the field squared near the boundary one gets%
\begin{equation}
\left\langle \varphi ^{2}\right\rangle _{\mathrm{c}}\approx -\frac{\Gamma
(D/2)(r-a)^{2-D}}{2^{D}(D-2)\pi ^{D/2+1}a}\int_{0}^{\infty }dx\,\frac{C(x)}{%
\left( 1+x^{2}\right) ^{(D-1)/2}}.  \label{phi2Near}
\end{equation}%
Note that for a spherical boundary in Minkowski spacetime on which the field
operator obeys Dirichlet or Neumann (or, in general, Robin) boundary
conditions, the VEV of the field squared diverges on the boundary as $%
(r-a)^{1-D}$ and the divergence is stronger.

\subsection{Vacuum energy-momentum tensor}

The VEV of the energy-momentum tensor is evaluated by using the formula (\ref%
{Tik}). The renormalization in the exterior region is reduced to the
subtraction of the part which corresponds to the Minkowski spacetime for all
$0\leqslant r<\infty $. The VEV of the energy-momentum tensor is diagonal.
For the renormalized components we get (no summation over $i$)%
\begin{equation}
\langle T_{i}^{i}\rangle _{\mathrm{c}}=\frac{\Gamma (D/2)}{4\pi ^{D/2+1}r^{D}%
}\sum_{l=0}^{\infty }D_{l}\int_{m}^{\infty }d\eta \,\eta \frac{\tilde{I}%
_{\nu _{l}}(a\eta )}{\tilde{K}_{\nu _{l}}(a\eta )}\frac{G_{\nu
_{l}}^{(i)}[K_{\nu _{l}}(r\eta )]}{\sqrt{\eta ^{2}-m^{2}}},  \label{Tll}
\end{equation}%
where for a given function $f(y)$ we define
\begin{eqnarray}
G_{\nu }^{(0)}[f(y)] &=&(4\xi -1)\left[ y^{2}f^{\prime 2}(y)-nyf(y)f^{\prime
}(y)+\left( \nu ^{2}-\frac{\left( 1+4\xi \right) y^{2}-2(mr)^{2}}{1-4\xi }%
\right) f^{2}(y)\right] ,  \notag \\
G_{\nu }^{(1)}[f(y)] &=&y^{2}f^{\prime 2}(y)+\xi _{1}yf(y)f^{\prime
}(y)-\left( y^{2}+\nu ^{2}+\xi _{1}n/2\right) f^{2}(y),  \label{Fnu1} \\
G_{\nu }^{(j)}[f(y)] &=&\left( 4\xi -1\right) y^{2}f^{\prime 2}(y)-\xi
_{1}yf(y)f^{\prime }(y)+\left[ \left( 4\xi -1\right) y^{2}+\frac{\nu
^{2}(1+\xi _{1})+\xi _{1}n/2}{n+1}\right] f^{2}(y),  \notag
\end{eqnarray}%
with $j=2,\ldots ,D$. In (\ref{Fnu1}), the notation
\begin{equation}
\xi _{1}=(D-1)\left( 4\xi -1\right) +1,  \label{ksi1}
\end{equation}%
is introduced. In general, the vacuum stresses along the radial and
azimuthal directions are isotropic.

It can be checked that the VEV given by (\ref{Tll}) obeys the covariant
conservation equation $\nabla _{k}\langle T_{i}^{k}\rangle _{\mathrm{c}}=0$
which, for the geometry under the consideration, is reduced to a single
equation%
\begin{equation}
r\partial _{r}\langle T_{1}^{1}\rangle _{\mathrm{c}}+(D-1)(\langle
T_{1}^{1}\rangle _{\mathrm{c}}-\langle T_{2}^{2}\rangle _{\mathrm{c}})=0.
\label{ConsEq}
\end{equation}%
We have also a trace relation%
\begin{equation}
\langle T_{i}^{i}\rangle _{\mathrm{c}}=\left[ D(\xi -\xi _{D})\nabla
_{l}\nabla ^{l}+m^{2}\right] \langle \varphi ^{2}\rangle _{\mathrm{c}}.
\label{TrRel}
\end{equation}%
In particular, the vacuum energy-momentum tensor is traceless for a
conformally coupled massless scalar field.

Now, let us investigate the behavior of the vacuum energy-momentum tensor at
large distances and near the boundary. At large distances from the sphere
and for a massive field, similarly to (\ref{phi2Large}), in the leading
order we get%
\begin{eqnarray}
\langle T_{0}^{0}\rangle _{\mathrm{c}} &\approx &\langle T_{2}^{2}\rangle _{%
\mathrm{c}}\approx -\frac{2mr}{D-1}\langle T_{1}^{1}\rangle _{\mathrm{c}}
\notag \\
&\approx &\frac{\Gamma (D/2)m^{2}(\xi -1/4)}{2\pi ^{(D-1)/2}r^{D-1}\sqrt{mr}%
e^{2mr}}\sum_{l=0}^{\infty }D_{l}\frac{\tilde{I}_{\nu _{l}}(am)}{\tilde{K}%
_{\nu _{l}}(am)}.  \label{T00Large}
\end{eqnarray}%
Note that in this region $|\langle T_{1}^{1}\rangle _{\mathrm{c}}|\ll
|\langle T_{0}^{0}\rangle _{\mathrm{c}}|$. For a massless field, assuming
that $r\gg a$, we introduce a new integration variable $y=r\eta $ in (\ref%
{Tll}) and expand the integrand over $a/r$. For a given $l$, the leading
term behaves like $(a/r)^{2l+2D-1}$ and it contains the integrals $%
\int_{0}^{\infty }dy\,y^{2\nu _{l}+2}F_{\nu _{l}}^{(i)}[K_{\nu _{l}}(y)]$.
These integrals are evaluated by using the relations%
\begin{eqnarray}
\int_{0}^{\infty }dy\,y^{2\nu +1}K_{\nu }(y)K_{\nu }^{\prime }(y) &=&-\left(
\nu +1/2\right) \mathcal{I}(\nu ),  \notag \\
\int_{0}^{\infty }dy\,y^{2\nu +2}K_{\nu }^{\prime 2}(y) &=&\left[ \nu
^{2}+\left( \nu +1/4\right) \frac{\nu +3/2}{\nu +1}\right] \mathcal{I}(\nu ),
\label{IntK2b}
\end{eqnarray}%
with the function $\mathcal{I}(\nu )$ defined by (\ref{IntK2}). These
relations are proved by making use of the well-known properties of the
Macdonald functions. The dominant contributions come from the terms with $l=0
$ and, to the leading order, for the energy density we find%
\begin{eqnarray}
\langle T_{0}^{0}\rangle _{\mathrm{c}} &\approx &-\frac{\left( \xi -\xi
_{D}\right) (a/r)^{2D-1}}{2^{D-2}\pi ^{D/2}a^{D+1}}\frac{D/2-1-\beta _{0}}{%
D/2-1+\beta _{0}}  \notag \\
&&\times \frac{\Gamma \left( D-1/2\right) \Gamma \left( (D+1)/2\right) }{%
\Gamma (D/2)\Gamma (D/2-1)}.  \label{T00Largem0}
\end{eqnarray}%
The asymptotics of the radial and azimuthal stresses are given by the
relations%
\begin{equation}
\langle T_{1}^{1}\rangle _{\mathrm{c}}\approx -\frac{D-1}{D}\langle
T_{2}^{2}\rangle _{\mathrm{c}}\approx -\langle T_{0}^{0}\rangle _{\mathrm{c}%
}.  \label{T11Largem0}
\end{equation}%
As it is seen, for a massless field, at large distances from the boundary
the radial pressure, $-\langle T_{1}^{1}\rangle _{\mathrm{c}}$, is equal to
the energy density. For a conformally coupled field the leading terms vanish.

The asymptotic behavior of the VEV of the energy-momentum tensor near the
boundary $r=a$ is investigated in a way similar to what we used for the
field squared. By using (\ref{IKtilde}) and the uniform asymptotic
expansions for the Macdonald function and its derivative, in the leading
order we obtain%
\begin{eqnarray}
\langle T_{0}^{0}\rangle _{\mathrm{c}} &\approx &\frac{(D-1)\Gamma (D/2)}{%
2^{D+2}\pi ^{D/2+1}a(r-a)^{D}}\int_{0}^{\infty }dx\,\frac{4\xi \left(
x^{2}+1\right) -1}{\left( x^{2}+1\right) ^{(D+1)/2}}C(x),  \notag \\
\langle T_{2}^{2}\rangle _{\mathrm{c}} &\approx &\frac{(D-1)\Gamma (D/2)}{%
2^{D+2}\pi ^{D/2+1}a(r-a)^{D}}\int_{0}^{\infty }dx\,\frac{C(x)}{\left(
1+x^{2}\right) ^{(D+1)/2}}  \notag \\
&&\times \left[ \left( 4\xi -1\right) \left( x^{2}+1\right) +\frac{1}{D-1}%
\right] ,  \label{T22Near}
\end{eqnarray}%
where the function $C(x)$, defined by (\ref{Cx}), depends on the specific
geometry in the region $r<a$. The leading term in the asymptotic expansion
of the radial stress is most easily found by making use of the continuity
equation (\ref{ConsEq}):%
\begin{equation}
\langle T_{1}^{1}\rangle _{\mathrm{c}}\approx -\frac{r-a}{a}\langle
T_{2}^{2}\rangle _{\mathrm{c}}.  \label{T11Near}
\end{equation}

For a spherical boundary in Minkowski spacetime with Dirichlet or Neumann
boundary conditions on the field operator the leading terms have the form
(no summation over $i$)%
\begin{equation}
\langle T_{i}^{i}\rangle _{\mathrm{c}}\approx \pm \frac{2D\Gamma ((D+1)/2)}{%
(4\pi )^{(D+1)/2}(r-a)^{D+1}}\left( \xi -\xi _{D}\right) ,  \label{TiiDN}
\end{equation}%
for $i=0,2,\ldots ,D$, and for the radial stress one has%
\begin{equation}
\langle T_{1}^{1}\rangle _{\mathrm{c}}\approx -\frac{D-1}{D}(r/a-1)\langle
T_{2}^{2}\rangle _{\mathrm{c}}.  \label{T11DN}
\end{equation}%
In (\ref{TiiDN}), the upper/lower sign corresponds to the Dirichlet/Neumann
boundary condition. The leading terms for the case of the Robin boundary
condition coincide with those for the Neumann condition. Similarly to the
case of the field squared, the divergences in these cases are stronger
compared to those for the geometry under consideration.

\section{Examples of the interior metric: dS and AdS spaces}

\label{sec:dS}

In this section, as examples of the interior metric we consider the
maximally symmetric spacetimes with positive and negative cosmological
constants, namely, dS and AdS spaces. As in the previous section the
exterior geometry is described by the Minkowski spacetime with the exterior
line element (\ref{ds2eM}). For the interior dS and AdS spaces the
corresponding line element in static coordinates has the form%
\begin{equation}
ds_{i}^{2}=(1+k\,\tilde{r}^{2}/\alpha ^{2})d\tilde{t}^{2}-(1+k\,\tilde{r}%
^{2}/\alpha ^{2})^{-1}d\tilde{r}^{2}-\tilde{r}^{2}d\Omega _{D-1}^{2},
\label{dSInt}
\end{equation}%
where $k=-1$ and $k=1$ for dS and AdS spaces, respectively. The parameter $%
\alpha $ is related to the cosmological constant $\Lambda $ through the
expression $\alpha =D(D-1)/(2|\Lambda |)$. In the case of dS space we assume
that the boundary is inside the dS horizon, corresponding to $\tilde{r}%
=\alpha $.

We should transform the line element in the form which is continuously glued
with the exterior Minkowskian line element at the boundary. To this aim,
firstly we introduce a new radial coordinate $r$ in accordance with
\begin{equation}
\tilde{r}=\alpha S_{k}(x),\;x=(r-r_{c})/\alpha ,  \label{rtilde}
\end{equation}%
where $r_{c}\leqslant r\leqslant a$ and%
\begin{equation}
S_{k}(x)=\left\{
\begin{array}{cc}
\sin x, & k=-1 \\
\sinh x, & k=1%
\end{array}%
\right. .  \label{Sk}
\end{equation}%
The parameter $r_{c}$ will be determined below by the matching conditions on
the boundary. Note that in the case of dS space the horizon corresponds to $%
x=\pi /2$ and, hence, to the value of the new radial coordinate $r=\pi
\alpha /2+r_{c}$. With the coordinate transformation (\ref{rtilde}), the
line element takes the form%
\begin{equation}
ds_{i}^{2}=C_{k}^{2}(x)d\tilde{t}^{2}-dr^{2}-\alpha ^{2}S_{k}^{2}(x)d\Omega
_{D-1}^{2},  \label{dsInt1}
\end{equation}%
where%
\begin{equation}
C_{k}(x)=\left\{
\begin{array}{cc}
\cos x, & k=-1 \\
\cosh x, & k=1%
\end{array}%
\right. .  \label{Ck}
\end{equation}%
Now the component $g_{11}$ of the metric tensor is continuous at $r=a$. From
(\ref{dsInt1}) it follows that the coordinate $r$ measures the proper
distance along the radial direction.

Next we define a new time coordinate $t$ in accordance with%
\begin{equation}
\tilde{t}=\frac{t}{C_{k}(x_{a})},\;x_{a}=\frac{a-r_{c}}{\alpha }.
\label{ttilde}
\end{equation}%
For the interior line element we get%
\begin{equation}
ds_{i}^{2}=\frac{C_{k}^{2}(x)}{C_{k}^{2}(x_{a})}dt^{2}-dr^{2}-\alpha
^{2}S_{k}^{2}(x)d\Omega _{D-1}^{2}.  \label{dsInt2}
\end{equation}%
In terms of the new coordinate $t$, at the boundary, the component $g_{00}$
is continuous as well. From (\ref{dsInt2}), for the interior functions in (%
\ref{metricinside}) one finds%
\begin{equation}
e^{u_{i}(r)}=\frac{C_{k}(x)}{C_{k}(x_{a})},\;v_{i}(r)=0,\;e^{w_{i}(r)}=%
\alpha S_{k}(x),  \label{metrInt2}
\end{equation}%
with $x$ given by (\ref{rtilde}).

The metric tensor corresponding to (\ref{dsInt2}) should be glued at $r=a$
with the exterior Minkowski spacetime in spherical coordinates with the line
element (\ref{ds2eM}) and with the functions (\ref{ueMink}). As we have
already noticed, the components $g_{00}$ and $g_{11}$ are continuous. From
the continuity of the components $g_{ll}$, $l=2,\ldots ,D$, we get%
\begin{equation}
S_{k}(x_{a})=a/\alpha ,  \label{Eqb}
\end{equation}%
with $x_{a}$ defined by (\ref{ttilde}). For given $a$ and $\alpha $ the
equation (\ref{Eqb}) determines the value of the parameter $r_{c}$. For dS
space, $r_{c}$ is negative ($1-\pi /2\leqslant r_{c}<0$ for $0\leqslant
a/\alpha <1$) and for AdS space it is positive. In the latter case and for
large values of $a/\alpha $ one has $r_{c}/\alpha \approx a/\alpha -\ln
(2a/\alpha )$. From (\ref{Eqb}) it follows that for the dS space the
boundary near the dS horizon ($x_{a}\rightarrow \pi /2$) corresponds to the
limit $a\rightarrow \alpha $. Note that, by taking into account (\ref{Eqb}),
for $C_{k}(x_{a})$ in (\ref{dsInt2}) one obtains
\begin{equation}
C_{k}(x_{a})=\sqrt{1+k(a/\alpha )^{2}}.  \label{Ckxa}
\end{equation}%
For the components of the surface energy-momentum tensor, from (\ref%
{matchcond2}), we get the expressions (no summation over $i$)%
\begin{eqnarray}
\tau _{0}^{0} &=&\frac{D-1}{8\pi Ga}[C_{k}(x_{a})-1],  \notag \\
\tau _{i}^{i} &=&\frac{D-2}{8\pi Ga}\left[ \frac{ka^{2}/\alpha ^{2}}{\left(
D-2\right) C_{k}(x_{a})}+C_{k}(x_{a})-1\right] ,  \label{taudS}
\end{eqnarray}%
with $i=2,\ldots ,D$. Note that the surface energy density is negative for
the interior dS space and is positive for the AdS space. In the case of dS
space, one has $C_{k}(x_{a})\rightarrow 0$ in the limit then the boundary
tends to the dS horizon, $a\rightarrow \alpha $. In this limit, the
azimuthal stress in (\ref{taudS}) diverges.

With the line element (\ref{dsInt2}), the equation (\ref{fleq}) for the
interior radial functions takes the form%
\begin{equation}
\frac{\partial _{x}\left[ C_{k}(x)S_{k}^{D-1}(x)\partial _{x}f_{(i)l}(r)%
\right] }{C_{k}(x)S_{k}^{D-1}(x)}+\left[ \frac{\alpha ^{2}\omega
^{2}C_{k}^{2}(x_{a})}{C_{k}^{2}(x)}-\frac{l(l+D-2)}{S_{k}^{2}(x)}-\alpha
^{2}\left( m^{2}+\xi R_{(i)}\right) \right] f_{(i)l}(r)=0,  \label{dSeq}
\end{equation}%
where the Ricci scalar is given by%
\begin{equation}
R_{(i)}=-kD(D+1)/\alpha ^{2}.  \label{RidS}
\end{equation}%
The solution of the equation (\ref{dSeq}), regular at the center, $x=0$, is
expressed in terms of the hypergeometric function as (see also \cite{Pola89})%
\begin{equation}
f_{(i)l}^{(1)}(r,\lambda )=\frac{[\tanh (\sqrt{k}x)/\sqrt{k}]^{l}}{\cosh
^{D/2+\nu }(\sqrt{k}x)}\,F(b_{l\lambda }^{(+)},b_{l\lambda
}^{(-)};l+D/2;\tanh ^{2}(\sqrt{k}x)),  \label{dSRegSol}
\end{equation}%
where we have introduced the notations%
\begin{eqnarray}
\nu  &=&\sqrt{D^{2}/4+k\alpha ^{2}m^{2}-D(D+1)\xi },  \notag \\
b_{l\lambda }^{(\pm )} &=&\frac{1}{2}[l+D/2+\nu \pm \sqrt{k}\alpha
C_{k}(x_{a})\sqrt{\lambda ^{2}+m^{2}}].  \label{blpm}
\end{eqnarray}%
For the dS interior the parameter $\nu $ can be either real or purely
imaginary. In the AdS case and for imaginary $\nu $ the ground state becomes
unstable \cite{Brei82,Mezi85}. By using formula 9.1.70 from \cite{Abra72},
it can be seen that in the limit $\alpha \rightarrow \infty $ the function $%
f_{(i)l}^{(1)}(r,\lambda )$ reduces to the function $r^{-n/2}J_{\nu
_{l}}(\lambda r)$, up to a constant coefficient. Note that in the
expressions of the VEVs in the exterior region the function (\ref{dSRegSol})
enters in the form (\ref{yi}) and, hence, the coefficient is not relevant.
The second linearly independent solution of (\ref{dSeq}) is given by the
expression%
\begin{equation}
f_{(i)l}^{(2)}(r,\lambda )=\frac{[\sqrt{k}\coth (\sqrt{k}x)]^{l+D-2}}{\cosh
^{D/2-\nu }(\sqrt{k}x)}\,F(1-b_{l\lambda }^{(+)},1-b_{l\lambda
}^{(-)};2-l-D/2;\tanh ^{2}(\sqrt{k}x)).  \label{dSIregSol}
\end{equation}%
By using the relation \cite{Abra72}
\begin{equation}
F(a,b;c;z)=(1-z)^{c-a-b}F(c-a,c-b;c;z),  \label{RelHyp}
\end{equation}%
for the hypergeometric function, we can see that the solutions (\ref%
{dSRegSol}) and (\ref{dSIregSol}) are symmetric under the change $\nu
\rightarrow -\nu $. In particular, from here it follows that these solutions
are real for purely imaginary values of $\nu $. We also have the property $%
f_{(i)l}^{(j)}(r,\lambda e^{\pi i})=f_{(i)l}^{(j)}(r,\lambda )$, $j=1,2$.

Now, the Wightman function and the VEVs of the field squared and of the
energy-momentum tensor in the exterior region are given by the equations (%
\ref{WMExt2}), (\ref{phi2}) and (\ref{Tll}), where now in the definition (%
\ref{Ftilde1}) one has%
\begin{equation}
\frac{8\pi G}{D-1}\tau a=\frac{ka^{2}/\alpha ^{2}}{C_{k}(x_{a})}+(D-1)\left[
C_{k}(x_{a})-1\right] .  \label{dStau}
\end{equation}%
In the expression of the logarithmic derivative of the radial function (\ref%
{dSRegSol}) we use the following formula for the derivative of the
hypergeometric function:%
\begin{equation}
(c-n)_{n}z^{c-1-n}F(a,b;c-n;z)=\partial _{z}^{n}\left[ z^{c-1}F(a,b;c;z)%
\right] ,  \label{HypDer}
\end{equation}%
where $(c)_{n}$ is Pochhammer's symbol. Taking $n=1$, we get%
\begin{equation}
\frac{\partial _{z}F(a,b;c;z)}{F(a,b;c;z)}=\frac{c-1}{z}\left[ \frac{%
F(a,b;c-1;z)}{F(a,b;c;z)}-1\right] .  \label{LogDer}
\end{equation}%
With the help of this formula, the expression of the logarithmic derivative
for the radial function (\ref{dSRegSol}) is presented in the form%
\begin{eqnarray}
&&\frac{\partial _{r}f_{(i)l}^{(1)}(r,\lambda )}{f_{(i)l}^{(1)}(r,\lambda )}=%
\frac{1}{\alpha \sqrt{kz}}\bigg\{l-\left( \nu _{l}+\nu +1\right) z  \notag \\
&&\qquad +2\left( 1-z\right) \nu _{l}\bigg[\frac{F(b_{l\lambda
}^{(+)},b_{l\lambda }^{(-)};\nu _{l};z)}{F(b_{l\lambda }^{(+)},b_{l\lambda
}^{(-)};\nu _{l}+1;z)}-1\bigg]\bigg\},  \label{LogDer2}
\end{eqnarray}%
with the notation%
\begin{equation}
z=\tanh ^{2}(\sqrt{k}x),  \label{z}
\end{equation}%
and with $x$ defined in (\ref{rtilde}).

In the expression of the VEVs in the exterior region, the logarithmic
derivative (\ref{LogDer2}) is evaluated at $r=a$. In this case
\begin{equation}
z|_{r=a}=z_{a}=\frac{1}{1+k\alpha ^{2}/a^{2}},  \label{za}
\end{equation}%
and in the notation (\ref{Ftilde1}) with tilde, one has%
\begin{eqnarray}
y_{l}(a,\eta ) &=&\frac{(kz_{a})^{-1/2}}{\alpha }\big\{l-\left( \nu _{l}+\nu
+1\right) z_{a}  \notag \\
&&+2\left( 1-z_{a}\right) \nu _{l}\left[ F_{\nu _{l}}(\eta ,z_{a})-1\right] %
\big\}.  \label{yla}
\end{eqnarray}%
Here, we have defined the function%
\begin{equation}
F_{\nu _{l}}(\eta ,z_{a})=\frac{F(b_{l}^{+}(\eta ),b_{l}^{-}(\eta );\nu
_{l};z_{a})}{F(b_{l}^{+}(\eta ),b_{l}^{-}(\eta );\nu _{l}+1;z_{a})},
\label{Fnul}
\end{equation}%
with%
\begin{equation}
b_{l}^{\pm }(\eta )=\frac{1}{2}\left[ \nu _{l}+\nu +1\pm i\sqrt{k}\alpha
C_{k}(x_{a})\sqrt{\eta ^{2}-m^{2}}\right] .  \label{blpmi}
\end{equation}%
and with $C_{k}(x_{a})$ given by (\ref{Ckxa}). Hence, for the interior dS
and AdS geometries, the VEVs of the field squared and the energy-momentum
tensor in the exterior Minkowskian region are given by (\ref{phi2}) and (\ref%
{Tll}), where in the expressions for $\tilde{I}_{\nu _{l}}(a\eta )$ and $%
\tilde{K}_{\nu _{l}}(a\eta )$, defined by (\ref{Ftilde1}), we should
substitute (\ref{dStau}) and (\ref{yla}).

The equation for bound states is obtained from (\ref{BoundStM}) with the
same substitutions. By a numerical calculation we have seen that, for a
given $a\eta $, the function $|\tilde{K}_{\nu _{l}}(a\eta )|$ increases with
increasing $l$ and, hence, if there are no bound states for $l=0$ the same
will be the case for higher $l$. For the interior AdS geometry the function $%
\tilde{K}_{\nu _{l}}(a\eta )$ is always negative and in this case there are
no bound states. For the dS interior the same is the case for a minimally
coupled field. The situation is changed in the case of dS interior geometry
for nonminimally coupled fields. We will discuss the features on the example
of a conformally coupled field.

\begin{figure}[tbph]
\begin{center}
\begin{tabular}{cc}
\epsfig{figure=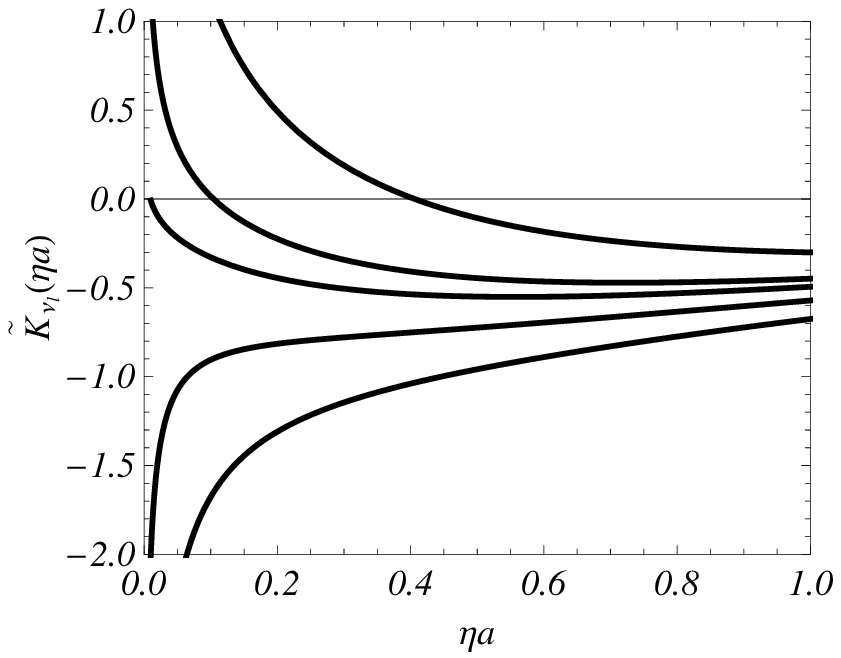,width=7.cm,height=6.cm} & \quad %
\epsfig{figure=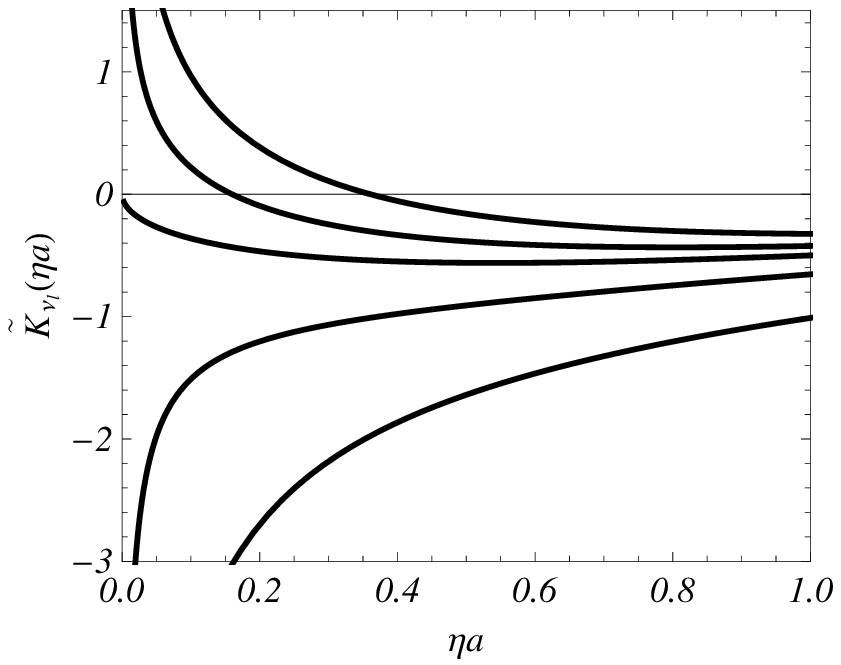,width=7.cm,height=6.cm}%
\end{tabular}%
\end{center}
\caption{The function $\tilde{K}_{\protect\nu _{l}}(a\protect\eta )$ in the
equation of the bound states for $l=0$ versus $a\protect\eta $ in the case
of a conformally coupled field in $D=3$ spatial dimensions. The left panel
corresponds to a massive field with $ma=1/4$ and the right panel presents
the case of a massless field. For the graphs on the left panel $\protect%
\alpha /a=1.0018,\,1.0025,\,1.00281,\,1.0035,\,1.005$ and for the right
panel $\protect\alpha /a=1.002,\,1.0025,\,1.00305,\,1.005,\,1.1$ (increasing
from top to bottom lines in the both cases). }
\label{fig1}
\end{figure}

If the dS horizon is not too close to the separating boundary, once again,
the function $\tilde{K}_{\nu _{l}}(a\eta )$ is negative and the bound states
are absent. However, bound states appear for $\alpha =\alpha _{1}>a$, where $%
\alpha _{1}$ is some critical value sufficiently close to $a$. With a
further decrease of $\alpha $, the value of $a\eta $ corresponding to the
bound state increases and, starting from the second critical value $\alpha
=\alpha _{2}$, it becomes larger than $ma$. This corresponds to the
imaginary value of the energy for the mode and signals the instability of
the exterior Minkowskian vacuum for $a<\alpha <\alpha _{2}$. For a massless
field, any possible real solution of the equation $\tilde{K}_{\nu
_{l}}(a\eta )=0$ leads to the instability of the exterior vacuum. We have
illustrated this type of situation for dS space in figure \ref{fig1}, where
for $l=0$ the function $\tilde{K}_{\nu _{l}}(a\eta )$ is plotted versus $%
a\eta $ for a conformally coupled scalar field in $D=3$ spatial dimensions.
For the left panel we have taken $ma=1/4$ and the right panel is for a
massless field. The curves on the left panel correspond to the values of the
ratio $\alpha /a=1.0018,\,1.0025,\,1.00281,\,1.0035,\,1.005$, increasing
from top to bottom lines. For the first critical value, corresponding to the
appearance of the bound state, one has $\alpha _{1}/a\approx 1.00281$. The
second critical value, starting from which the vacuum becomes unstable,
corresponds to $\alpha _{2}/a\approx 1.0021$. The left zero on the left
panel corresponds to a bound state ($\eta _{s}<m$), whereas the right zero
corresponds to an unstable mode ($\eta _{s}>m$). For the curves on the right
panel we have $\alpha /a=1.002,\,1.0025,\,1.00305,\,1.005,\,1.1$ (increasing
from top to bottom lines). Here, any solution of the equation (\ref{BoundStM}%
) corresponds to the instability and for the critical value of the dS
curvature radius we have $\alpha _{2}/a\approx 1.00305$.

Now we turn to the investigation of the asymptotic behavior of the VEVs. At
large distances, the asymptotics are given by (\ref{phi2Large}), (\ref%
{T00Large}) and (\ref{phi2Largem0}), (\ref{T00Largem0}) for massive and
massless fields, respectively. For the interior geometries under
consideration, the quantity $\beta _{0}$, appearing in the asymptotic for a
massless field, is given by the expression%
\begin{eqnarray}
\beta _{0} &=&\frac{1}{C_{k}(x_{a})}\left\{ 2k\left( a/\alpha \right)
^{2}\left( \xi -b_{0}\right) +n\left[ F(z_{a})-1\right] \right\}   \notag \\
&&+2\xi (D-1)\left[ C_{k}(x_{a})-1\right] +n/2,  \label{bet0b}
\end{eqnarray}%
where $b_{0}=D/4+\nu /2$ and we have defined the function%
\begin{equation}
F(z_{a})=\frac{F(b_{0},b_{0};D/2-1;z_{a})}{F(b_{0},b_{0};D/2;z_{a})}.
\label{FzLarge}
\end{equation}%
For a minimally coupled field, for this function one has%
\begin{equation}
F(z_{a})=1+\frac{kD}{D-2}\frac{a^{2}}{\alpha ^{2}},  \label{Fzmin}
\end{equation}%
and from (\ref{bet0b}) we get $\beta _{0}=D/2-1$. Now, from (\ref%
{phi2Largem0}) and (\ref{T00Largem0}) we see that the leading terms in the
asymptotic expansion of the VEVs at large distances vanish and the decay for
this case is stronger. For a conformally coupled field and for $D=3$ one has
$b_{0}=1$ and the function $F(z)$ in (\ref{FzLarge}) is reduced to%
\begin{equation}
F(z_{a})=k\frac{a^{2}}{\alpha ^{2}}+\frac{a/\alpha }{A_{k}(1/\sqrt{\alpha
^{2}/a^{2}+k})},  \label{Fzcc}
\end{equation}%
with%
\begin{equation}
A_{k}(x)=\left\{
\begin{array}{cc}
\mathrm{arcsinh}\,x, & k=-1 \\
\arcsin x, & k=1%
\end{array}%
\right. .  \label{Ak}
\end{equation}%
With the function (\ref{Fzcc}) in (\ref{bet0b}) one has $(n-2\beta
_{0})/(n+2\beta _{0})>0$ for all values of $a/\alpha $ in the AdS case and
for $a/\alpha <a/\alpha _{2}$ for the dS interior. In the latter case $%
a/\alpha _{2}\approx 0.997$ is the critical value for the vacuum instability
(see the right panel in figure \ref{fig1}). Now, from (\ref{phi2Largem0}) it
follows that the corresponding VEV of the field squared is positive at large
distances.

In order to find the leading terms in the asymptotic expansions of the VEVs
near the boundary by using the general formulae (\ref{phi2Near}) and (\ref%
{T22Near}), we need the function $B(x)$ in the asymptotic expansion (\ref%
{yia}) for the interior spaces under consideration. This function is found
in Appendix \ref{sec:Append}. By using the expression for the function $C(x)$
from (\ref{Cu}), the integral in (\ref{phi2Near}) is expressed in terms of
the gamma functions and for the VEV of the field squared one gets%
\begin{equation}
\left\langle \varphi ^{2}\right\rangle _{\mathrm{c}}\approx -\frac{\left(
\xi -\xi _{D}\right) \Gamma ((D-1)/2)}{2^{D}\pi ^{(D+1)/2}(D-2)a}\frac{%
C_{k}(x_{a})-1}{(r-a)^{D-2}}\left[ D+\frac{1}{C_{k}(x_{a})}\right] .
\label{phi2NeardS}
\end{equation}%
This leading term does not depend on the mass of the field. For a
conformally coupled field it vanishes and the next-to-leading order term
should be kept. For a minimally coupled field, near the boundary the VEV of
the field squared is negative for the interior dS space and positive for the
AdS space.

\begin{figure}[tbph]
\begin{center}
\begin{tabular}{cc}
\epsfig{figure=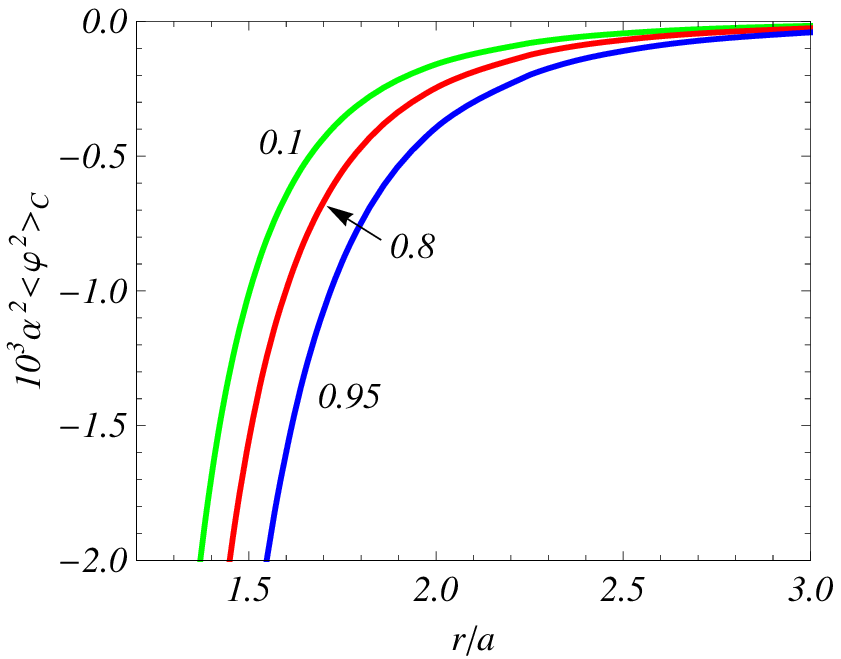,width=7.cm,height=6.cm} & \quad %
\epsfig{figure=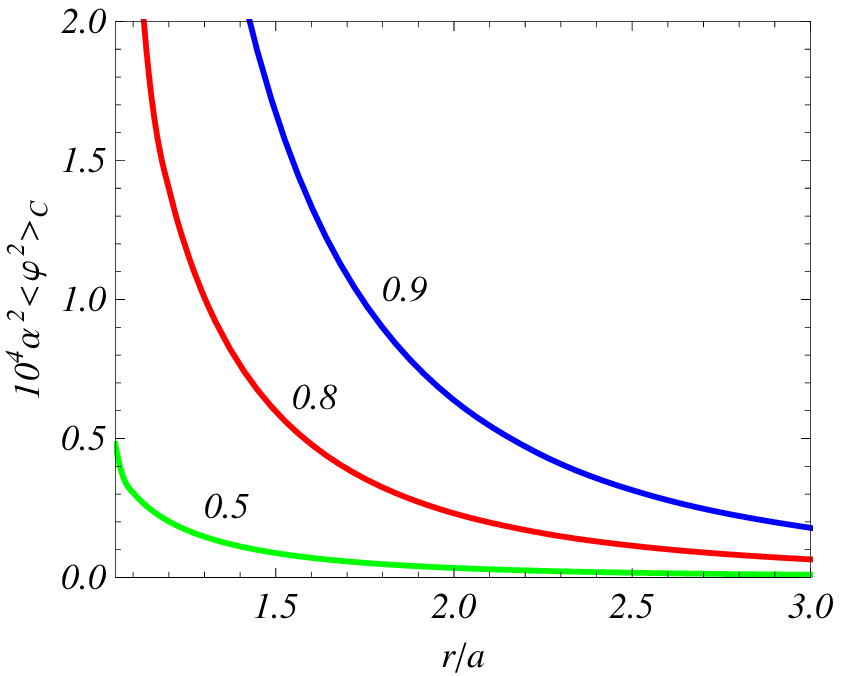,width=7.cm,height=6.cm}%
\end{tabular}%
\end{center}
\caption{VEV of the field squared, $\protect\alpha ^{D-1}\left\langle
\protect\varphi ^{2}\right\rangle _{\mathrm{c}}$, for the interior $D=3$ dS
geometry, as a function of the rescaled radial coordinate for several values
of $a/\protect\alpha $ (numbers near the curves). The left and right panels
correspond to minimally and conformally coupled massless scalar fields.}
\label{fig2}
\end{figure}

\begin{figure}[tbph]
\begin{center}
\begin{tabular}{cc}
\epsfig{figure=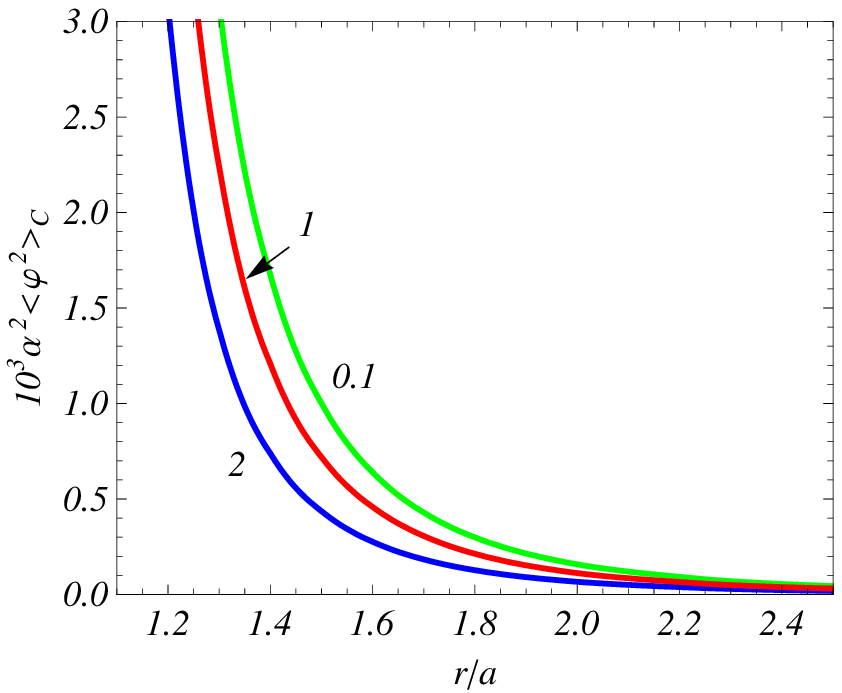,width=7.cm,height=6.cm} & \quad %
\epsfig{figure=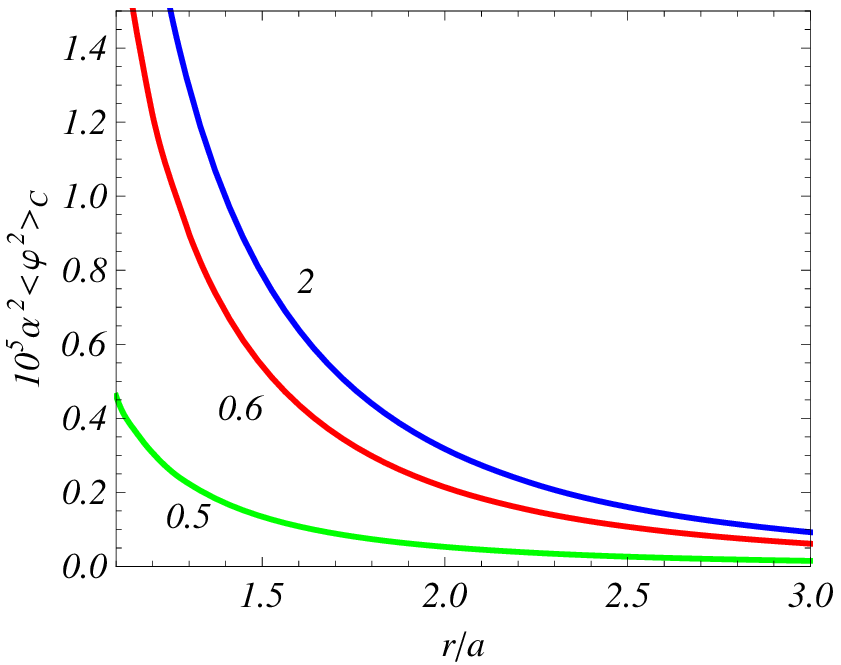,width=7.cm,height=6.cm}%
\end{tabular}%
\end{center}
\caption{The same as in figure \protect\ref{fig2} in the case of the
interior AdS space.}
\label{fig3}
\end{figure}

In figure \ref{fig2}, for the dS interior geometry, we have plotted the VEV
of the field squared in the exterior region, $\alpha ^{D-1}\left\langle
\varphi ^{2}\right\rangle _{\mathrm{c}}$, in $D=3$ spatial dimensions, as a
function of the ratio $r/a$. The numbers near the curves are the values of $%
a/\alpha $. The left and right panels correspond to minimally ($\xi =0$) and
conformally ($\xi =1/6$) coupled massless scalars. Similar graphs for the
AdS interior geometry are presented in figure \ref{fig3}.

In a similar way, from (\ref{T22Near}) for the VEVs of the energy density
and the azimuthal stress near the boundary we obtain%
\begin{eqnarray}
\langle T_{0}^{0}\rangle _{\mathrm{c}} &\approx &\frac{\Gamma ((D+1)/2)\left[
C_{k}(x_{a})-1\right] }{2^{D-1}\pi ^{(D+1)/2}a(r-a)^{D}}  \notag \\
&&\times \left\{ \,\left( \xi -\xi _{D}\right) \left[ D\left( \xi -\xi
_{D}\right) +\frac{\xi }{C_{k}(x_{a})}\right] -\xi _{D}\frac{\left( \xi -\xi
_{D+2}\right) }{C_{k}(x_{a})}\right\} .  \label{T00NeardS}
\end{eqnarray}%
and%
\begin{eqnarray}
\langle T_{2}^{2}\rangle _{\mathrm{c}} &\approx &\frac{\Gamma ((D+1)/2)\left[
C_{k}(x_{a})-1\right] }{2^{D-1}\pi ^{(D+1)/2}a(r-a)^{D}}  \notag \\
&&\times \left\{ \left( \xi -\xi _{D}\right) \left[ D\left( \xi -\xi
_{D}\right) +\frac{\xi -1/4}{C_{k}(x_{a})}\right] +\frac{\xi -\xi _{D+2}}{%
4DC_{k}(x_{a})}\right\} ,  \label{T22NeardS}
\end{eqnarray}%
where $\xi _{D+2}=(D+1)/(4(D+2))$. The leading term in the asymptotic
expansion of the radial stress is found by using the relation (\ref{T11Near}%
). For a minimally coupled field, the energy density, $\langle
T_{0}^{0}\rangle _{\mathrm{c}}$, and azimuthal stress, $\langle
T_{2}^{2}\rangle _{\mathrm{c}}$, are negative near the boundary for the dS
interior space and are positive for the AdS interior. The expressions (\ref%
{T00NeardS}) and (\ref{T22NeardS}) are further simplified for a conformally
coupled field%
\begin{equation}
\langle T_{2}^{2}\rangle _{\mathrm{c}}\approx -\frac{1}{D-1}\langle
T_{0}^{0}\rangle \approx \frac{\Gamma ((D+1)/2)\left[ 1/C_{k}(x_{a})-1\right]
}{2^{D+3}\pi ^{(D+1)/2}D^{2}(D+2)a(r-a)^{D}}.  \label{T22NeardSCP}
\end{equation}%
In this case, near the boundary the vacuum energy and the azimuthal pressure
($-\langle T_{2}^{2}\rangle _{\mathrm{c}}$) are negative for the interior dS
space and are positive for the AdS space.

\begin{figure}[tbph]
\begin{center}
\begin{tabular}{cc}
\epsfig{figure=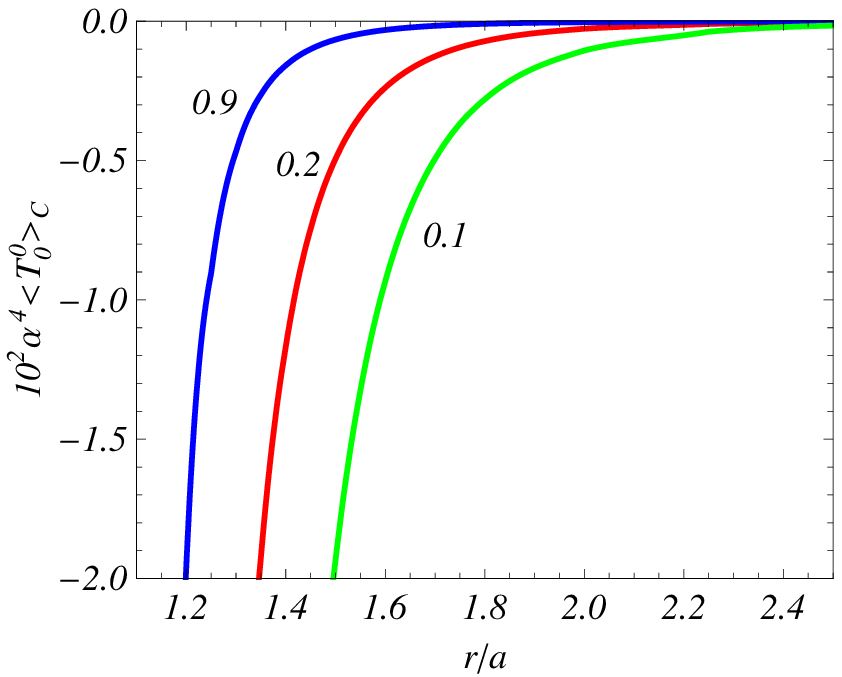,width=7.cm,height=6.cm} & \quad %
\epsfig{figure=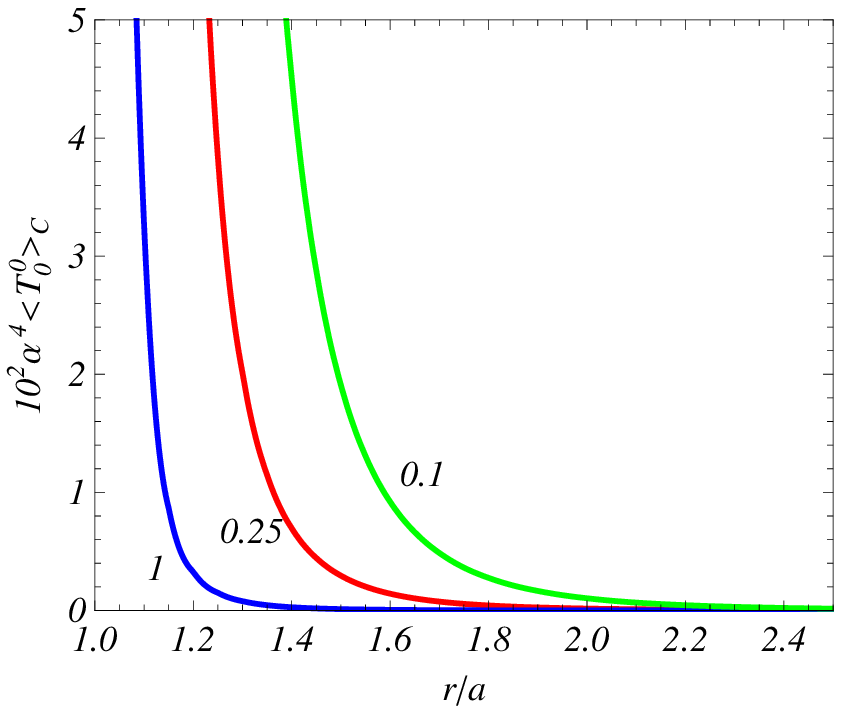,width=7.cm,height=6.cm}%
\end{tabular}%
\end{center}
\caption{VEV of the energy density, $\protect\alpha ^{D+1}\langle
T_{0}^{0}\rangle _{\mathrm{c}}$ for a conformally coupled massless field,
induced by the interior $D=3$ dS space (left panel) and AdS (right panel)
spaces, versus $r/a$. The numbers near the curves correspond to the values
of the ratio $a/\protect\alpha $. }
\label{fig4}
\end{figure}

Figure \ref{fig4} displays the VEV of the energy density, $\alpha
^{D+1}\langle T_{0}^{0}\rangle _{\mathrm{c}}$, induced by the interior $D=3$
dS (left panel) and AdS (right panel) geometries, for a conformally coupled
massless scalar field, as a function of the rescaled radial coordinate. The
numbers near the curves correspond to the values of the parameter $a/\alpha $%
. The corresponding graphs for a massless minimally coupled scalar field
show similar behavior.

\begin{figure}[tbph]
\begin{center}
\begin{tabular}{cc}
\epsfig{figure=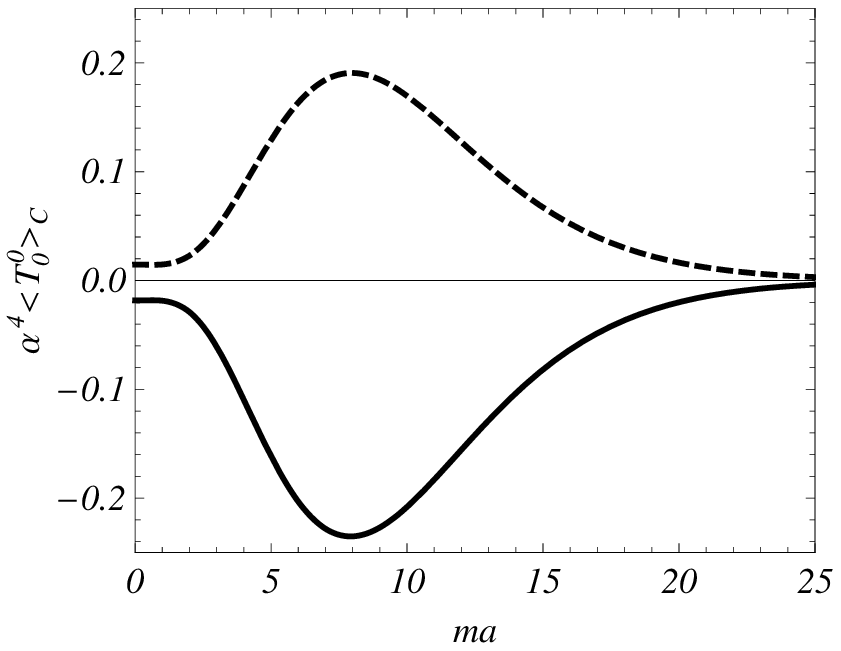,width=7.cm,height=6.cm} & \quad %
\epsfig{figure=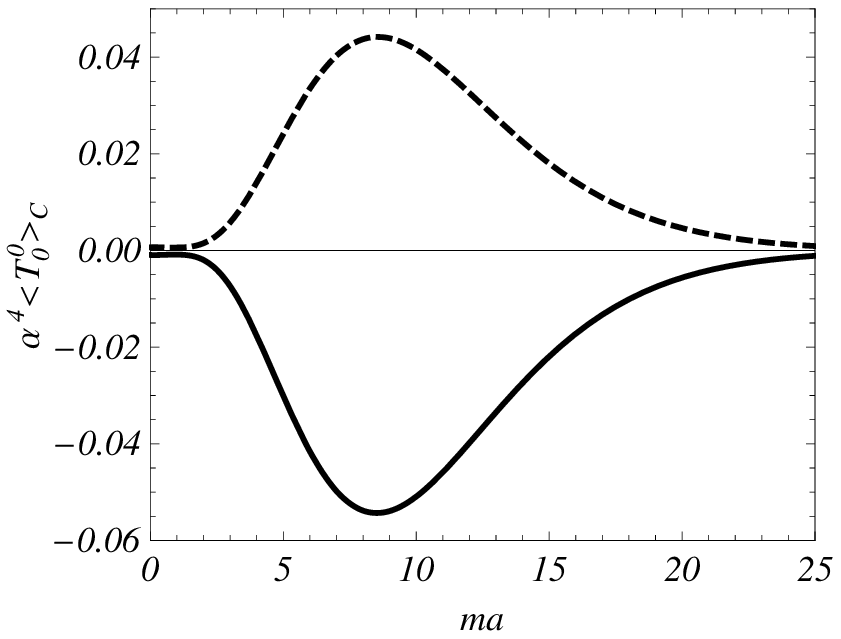,width=7.cm,height=6.cm}%
\end{tabular}%
\end{center}
\caption{VEV of the energy density in the exterior region as a function of $m%
\protect\alpha $ for fixed $a/\protect\alpha =0.5$ and $r/a=1.5$. The left
and right panels correspond to minimally and conformally coupled scalar
fields in $D=3$ spatial dimensions and the full and dashed curves correspond
to dS and AdS interiors.}
\label{fig5}
\end{figure}

It is also of interest to consider the dependence of the VEVs on the mass of
the field. In figure \ref{fig5} we have plotted the VEV of the energy
density in the exterior region as a function of $m\alpha $, for fixed values
$a/\alpha =0.5$, $r/a=1.5$, in the cases of minimally (left panel) and
conformally (right panel) coupled fields in $D=3$ spatial dimensions. The
full and dashed curves correspond to interior dS and AdS spaces. As it is
seen from the graphs, the VEV is not a monotonic function of the mass.

In the investigation of the VEVs for the case of the interior dS space we
have assumed that $\alpha /a>1$. For the interior AdS space the value of
this ratio can be arbitrary. In this case, it is of interest to consider the
behavior of the VEVs for small values of the AdS curvature radius, $\alpha
/a\ll 1$, corresponding to a strong gravitational field in the interior
region. In this limit, the argument of the hypergeometric functions in (\ref%
{Fnul}) is close to 1, $\left( 1-z_{a}\right) \ll 1$. By using the formula
15.3.6 from \cite{Abra72}, to the leading order we get%
\begin{equation}
F_{\nu _{l}}(\eta ,z_{a})\approx \left( \alpha /a\right) ^{-2}\nu /\nu _{l}.
\end{equation}%
The coefficient of the function $F(z)$ in (\ref{Ftilde1}) becomes $(a/\alpha
)\nu ^{\prime }$ with the notation $\nu ^{\prime }=\nu +2\xi D-D/2$. For $%
\nu ^{\prime }\neq 0$, in the limit under consideration, this coefficient is
large and, in the leading order, the VEVs in the exterior region coincide
with the corresponding VEVs for a spherical boundary in Minkowski spacetime
with the Dirichlet boundary condition. For $\nu ^{\prime }=0$ the
next-to-leading term in the expansion over $\alpha /a$ should be taken into
account. Keeping this term, we can see that the VEVs are reduced to those
for a spherical shell with Dirichlet and Neumann boundary conditions in the
cases $\nu <1/2$ and $\nu >1/2$, respectively. The case $\nu ^{\prime }=0$
with $\nu =1/2$ corresponds to a conformally coupled massless scalar field
and in this case the VEVs are not reduced to Dirichlet or Neumann results.
If, in addition, we assume that $\alpha m\ll 1$, then the condition $\nu
^{\prime }=0$ is satisfied for the special cases of minimally and
conformally coupled fields. We expect that, for small values of the AdS
curvature radius, the VEVs in the interior region will be suppressed. This
sort of suppression in the boundary-induced local VEVs for the geometry of
parallel plates in AdS bulk, described in Poincar\'{e} coordinates, has been
discussed in \cite{Saha05,Saha06,Eliz13} for scalar and fermionic fields.

We have considered the VEVs in a combined geometry with interior dS or AdS
and exterior Minkowski spacetimes. It would be interesting to generalize the
corresponding results for the exterior Schwarzschild solution of the
Einstein equations. The possibility that the interior geometry of a black
hole could be constituted by a dS region has been discussed in the
literature (see \cite{Blau87}-\cite{Dymn05} and references therein).
However, in the Schwarzshild geometry the equation for the radial part of
the scalar mode functions is not exactly solvable and numerical or
approximate results only can be provided.

\section{Conclusion}

\label{sec:Conc}

In the present paper we have considered the Casimir densities for a scalar
field with a general curvature coupling parameter, induced by a spherical
boundary separating the spacetime backgrounds with different geometries. The
latter are described by spherically symmetric static line elements (\ref%
{metricinside}) and (\ref{metricoutside}) for the interior and exterior
regions respectively. Additionally, the presence of an infinitely thin
spherical shell with a surface energy-momentum tensor $\tau _{i}^{k}$ is
assumed. The interior and exterior metric tensors are continuous on the
separating boundary and their radial derivatives are related by the Israel
matching conditions. The latter lead to the relations (\ref{matchcond2}) for
the functions in the expressions of the metric tensor components. The
matching conditions for a scalar field are obtained from the corresponding
field equation: the field is continuous on the separating surface and the
jump in the radial derivative is given by the relation (\ref{DerJump}). The
jump comes from the nonminimal coupling of the field and is a consequence of
the delta function term in the Ricci scalar located on the separating
boundary.

For the investigation of the exterior vacuum properties induced by the
interior geometry, first we evaluate the positive frequency Wightman
function with the help of the direct summation over a complete set of field
modes. In Section \ref{sec:Modes}, for the general cases of interior and
exterior geometries, we have constructed a complete set of normalized mode
functions obeying the matching conditions. In addition to the modes with
continuous energy spectrum, depending on background geometry, the modes
describing the bound states can be present. For these modes the quantum
number $\lambda $ is purely imaginary and the corresponding eigenvalues for $%
\eta =|\lambda |$ are solutions of the equation (\ref{BoundSt}) with the
notation (\ref{Fhat}). The Wightman function in the exterior region is given
by the expression (\ref{WFext}) for the modes with continuous energy
spectrum and by (\ref{Wbs}) for the contribution coming from the bound
states. In order to separate from the expression of the Wightman function
the part induced by the interior geometry, we use the identity (\ref{Ident}%
). Then, by using the asymptotic properties of the radial parts in the mode
functions, we rotate the contours of the integration in the complex plane $%
\lambda $. As a result, the Wightman function in the exterior region is
presented in a decomposed form (\ref{Wdec2}). In this representation, the
function $W_{0}(x,x^{\prime })$ is the Wightman function in the case of the
background described by the line element (\ref{metricoutside}) for all
values of the radial coordinate $r$ and the contribution $W_{\mathrm{c}%
}(x,x^{\prime })$ is induced by the geometry (\ref{metricinside}) in the
region $r<a$. Compared with the initial form, the representation (\ref{Wdec2}%
) of the Wightman function has two important advantages. First of all, in
the part induced by the interior geometry the integrand is an exponentially
decreasing function at the upper limit of the integration, instead of highly
oscillatory behavior in the initial representation. And, secondly, for
points outside the boundary, the divergences arising in the coincidence
limit of the arguments are contained in the part $W_{0}(x,x^{\prime })$,
whereas the part induced by the interior geometry is finite in the
coincidence limit. With this property, the renormalization of the VEVs for
the field squared and the energy-momentum tensor is reduced to the
renormalization for the background (\ref{metricoutside}) for all values of $r
$. Hence, the contributions to the VEVs coming from the interior geometry
are directly obtained from the corresponding part of the Wightman function
without any additional subtractions.

For a given Wightman function, the VEVs of the field squared and the
energy-momentum tensor are evaluated by formulae (\ref{Tik}). They are
decomposed as (\ref{Tikdec}), where the second terms in the right-hand sides
are induced by the geometry (\ref{metricinside}) in the region $r<a$. These
terms are obtained from the corresponding part in the Wightman function
without additional renormalization. For example, the VEV of the field
squared is given by (\ref{phi2Gen}).

A special case, with the Minkowski spacetime as an exterior geometry, is
discussed in Section \ref{sec:ExtMink}. In this case the expression for the
Wightman function in the exterior region is reduced to (\ref{WMExt2}). The
latter differs from the corresponding expression for a spherical boundary
with Robin boundary condition by the replacement (\ref{Rob}) of the Robin
coefficient. In the geometry under consideration, the 'effective' Robin
coefficient depends on the quantum numbers specifying the scalar field modes
and this leads to important modifications in the behavior of the VEVs near
the boundary. For the exterior Minkowskian geometry, the VEVs of the field
squared and the energy-momentum tensor are given by the expressions (\ref%
{phi2}) and (\ref{Tll}). For a massive field, at distances from the boundary
larger than the Compton wavelength, the VEVs are exponentially suppressed.
For a massless field the decay of the VEVs at large distances is power-law:
it goes like $r^{3-2D}$ for the field squared and like $r^{1-2D}$ for the
energy-momentum tensor. The exponents in the power-law decay are different
in the special cases $\beta _{0}=\pm (D/2-1)$ with $\beta _{0}$ defined by (%
\ref{bet0}). The VEVs diverge on the boundary separating the interior and
exterior geometries. The leading terms in the asymptotic expansions over the
distance from the boundary are given by (\ref{phi2Near}) for the field
squared and by (\ref{T22Near}) for the energy density and the azimuthal
stress. For the radial stress near the boundary one has (\ref{T11Near}). The
function $C(x)$ in the expressions for the leading terms is determined from
the uniform asymptotic expansion of the interior radial mode function for
large values of the orbital momentum and it depends on the specific interior
geometry. The VEV of the field squared diverges on the boundary as $%
(r/a-1)^{2-D}$ and the VEVs of the energy-density and the azimuthal stress
diverge as $(r/a-1)^{-D}$. In the case of a spherical boundary in Minkowski
spacetime with Dirichlet and Neumann (or, in general, Robin) boundary
conditions the surface divergences are stronger.

As an application of general results, in Section \ref{sec:dS} we have
considered dS and AdS spaces as examples of the interior geometry. Firstly
we have transformed the corresponding line elements to the form (\ref{dsInt2}%
) which is continuously matched with the exterior Minkowskian geometry. The
components of the corresponding surface energy-momentum tensor are given by (%
\ref{taudS}). The radial parts of the interior mode functions are expressed
in terms of the hypergeometric function ((\ref{dSRegSol}) and (\ref%
{dSIregSol}) for regular and irregular modes, respectively). The VEVs of the
field squared and the energy-momentum tensor in the exterior Minkowskian
region are determined by the formulae (\ref{phi2}) and (\ref{Tll}), where in
the expressions for $\tilde{I}_{\nu _{l}}(a\eta )$ and $\tilde{K}_{\nu
_{l}}(a\eta )$, defined by (\ref{Ftilde1}), the functions (\ref{dStau}) and (%
\ref{yla}) should be substituted. In the case of the interior AdS geometry
there are no bound states. For the dS interior the same holds for a
minimally coupled field. In the case of the dS interior geometry and for
nonminimally coupled fields, bound states are absent if the radius of the
separating boundary is not too close to the dS horizon radius. When the
boundary becomes closer to the horizon, bound states appear. With the
further increasing of the boundary radius, the energy of the bound state
decreases, and for some critical value it becomes zero. The further increase
leads to imaginary values of the energy thus signaling the exterior
Minkowski vacuum instability.

In the cases of dS and AdS interior spaces, we have specified the general
formulae for the asymptotics of the VEVs. The parameter $\beta _{0}$,
determining the large distance behavior of the VEVs for massless fields, is
given by the expression (\ref{bet0b}). For a minimally coupled field one has
$\beta _{0}=D/2-1$ and the leading terms in the asymptotic expansion of the
VEVs at large distances vanish. The leading terms in the expansions near the
boundary are given by the expressions (\ref{phi2NeardS}), (\ref{T00NeardS})
and (\ref{T22NeardS}). For a conformally coupled field the leading term in
the VEV of the field squared vanishes. In this case, near the boundary the
vacuum energy and the azimuthal pressure are negative for the interior dS
space and are positive for the AdS space. For a minimally coupled field and
near the boundary, the VEVs of the field squared, energy density and
azimuthal stress are negative for the interior dS space and positive for the
AdS space. In the latter case and for small values of the AdS curvature
radius (strong gravitational field in the interior region), $\alpha \ll
a,m^{-1}$, for the curvature coupling parameter $\xi \neq 0,\xi _{D}$, the
VEVs in the exterior region, to the leading order, coincide with the
corresponding VEVs for a spherical boundary in Minkowski spacetime with
Dirichlet boundary condition. For a minimally coupled field, the VEVs are
reduced to those for a spherical shell with Neumann boundary condition. In
the special case of the conformal coupling, the VEVs are not reduced to
Dirichlet or Neumann results.

The results given above for gravitational backgrounds may have applications
in effective field theoretical models of some condensed matter systems
formulated on curved backgrounds (see, for example, \cite{Klin05,Volo03}).
An important example of this sort are graphene-made structures. The
long-wavelength description of the graphene excitations can be formulated in
terms of the effective field theory in $(2+1)$-dimensional spacetime. In the
geometry of a single-walled carbon nanotube, which is generated by rolling
up a graphene sheet to form a cylinder, the background space is flat and has
topology $R^{1}\times S^{1}$. For nanotubes with open ends, the Casimir
densities induced by the nontrivial topology and by the edges have been
discussed in \cite{Bell09,Eliz11,Bell13}. However, the end of the nanotube
can be closed with a hemispherical cap. In this case the geometry for the
corresponding effective field theory is of the type discussed above with the
interior constant curvature space.

\section*{Acknowledgments}

AAS was supported by State Committee Science MES RA, within the frame of the
research project No. SCS 13-1C040.

\appendix

\section{Asymptotic of the hypergeometric function}

\label{sec:Append}

As it has been shown in Section \ref{sec:ExtMink}, the leading terms of the
asymptotic expansions for the VEVs near the spherical boundary, separating
the regions with different geometries, are expressed in terms of the
function $C(x)$ given by (\ref{Cx}). In this expression, $B(x)$ is defined
by the asymptotic expansion of the function $y_{l}(a,\nu _{l}\eta )$ for
large $\nu _{l}$ (see (\ref{yia})). In order to find the function $B(x)$ for
the special cases of the interior geometry corresponding to dS and AdS
spaces, in accordance with (\ref{LogDer2}), we need the asymptotic of the
function $F_{\nu _{l}}(\nu _{l}\lambda ,z_{a})$ for large values of $\nu _{l}
$. The leading term is obtained from the general consideration given above,
and for the determination of the function $B(x)$ we need the next-to-leading
term. In the limit under consideration, all the parameters of the
hypergeometric functions in (\ref{Fnul}) are large. The corresponding
asymptotics have been recently investigated in \cite{Pari13a,Pari13b}. By
using the expansion (2.8) from \cite{Pari13b}, for large $|\mu |$ the
following result can be obtained:%
\begin{equation}
\frac{F(a+\varepsilon _{1}\mu ,b+\varepsilon _{2}\mu ;c+\mu -1;z)}{%
F(a+\varepsilon _{1}\mu ,b+\varepsilon _{2}\mu ;c+\mu ;z)}\sim \frac{%
1-\varepsilon _{1}}{1-t_{s}}\left[ 1+\frac{h(t_{s})}{\mu }+\cdots \right] ,
\label{Hypas}
\end{equation}%
where $0<\varepsilon _{1}\leqslant \varepsilon _{2}<1$ and (in notations of
\cite{Pari13b})%
\begin{equation}
t_{s}=\frac{\Delta -\sqrt{\Delta ^{2}-4\varepsilon _{1}(1-\varepsilon _{2})z}%
}{2(1-\varepsilon _{2})z},\;\Delta =1+\left( \varepsilon _{1}-\varepsilon
_{2}\right) z.  \label{ts}
\end{equation}%
In (\ref{Hypas}), we have defined the function%
\begin{eqnarray}
h(t) &=&\frac{\left( c-1\right) \varepsilon _{1}-a}{1-\varepsilon _{1}}-%
\frac{t}{h_{1}(t)}\Big\{\left[ \left( c-3\right) \varepsilon _{2}-b\right]
t+b\varepsilon _{1}-\varepsilon _{2}\left( a-1\right)   \notag \\
&&+\frac{1}{h_{1}(t)}\left[ \left( \varepsilon _{2}^{2}-1\right) \left(
t-3\varepsilon _{1}\right) t^{2}+3\varepsilon _{1}\left( \varepsilon
_{2}^{2}-\varepsilon _{1}\right) t+\varepsilon _{1}\left( \varepsilon
_{1}^{2}-\varepsilon _{2}^{2}\right) \right] \Big\},  \label{ht}
\end{eqnarray}%
with%
\begin{equation}
h_{1}(t)=\varepsilon _{1}(\varepsilon _{2}-\varepsilon _{1})+t\left(
2\varepsilon _{1}-t\right) (1-\varepsilon _{2}).  \label{h1t}
\end{equation}

In order to apply (\ref{Hypas}) to the function $F_{\nu _{l}}(\nu _{l}\eta
,z_{a})$ (defined by (\ref{Fnul})) with $\mu =\nu _{l}$, we assume for the
moment that $k=-1$. In this case the parameters $\varepsilon _{j}$
corresponding to (\ref{Hypas}) are real. We are interested in the term of
the order $1/\nu _{l}$, and to this order, the mass term in (\ref{blpmi})
does not contribute. Assuming that the parameters are in the range required
for the validity of (\ref{Hypas}), we take in this expansion%
\begin{eqnarray}
&& a=b=\frac{1}{2}\left( 1+\nu \right) ,\;c=1,\;  \notag \\
&& \varepsilon _{1} =\frac{1}{2}\left( 1-\gamma \right) ,\;\varepsilon _{2}=%
\frac{1}{2}\left( 1+\gamma \right) ,  \label{eps12}
\end{eqnarray}%
with%
\begin{equation}
\gamma =\eta \sqrt{\alpha ^{2}-a^{2}}.  \label{alfas}
\end{equation}%
For these values of the parameters one has%
\begin{equation}
2\frac{1-\varepsilon _{1}}{1-t_{s}}=1+\sqrt{1-a^{2}/\alpha ^{2}}\sqrt{%
1+a^{2}\eta ^{2}}.  \label{LeadTerm}
\end{equation}%
In the leading order this gives%
\begin{equation}
2F_{\nu _{l}}(\nu _{l}\eta ,z_{a})\sim 1+\sqrt{1-a^{2}/\alpha ^{2}}\sqrt{%
1+a^{2}\eta ^{2}}.  \label{Fleading}
\end{equation}%
Substituting into (\ref{yla}), we obtain the leading term for the expansion
of the function $y_{l}(a,\nu _{l}\lambda )$ which agrees with the result (%
\ref{yia}) obtained directly from the differential equation for $%
y_{l}(a,\eta )$.

Evaluating the function $h(t)$ for special values of the parameters (\ref%
{eps12}), (\ref{alfas}) and substituting the corresponding expansion (\ref%
{Hypas}) into the expression (\ref{yla}) with $\eta \rightarrow \nu _{l}\eta
$, after long calculations we find the expansion (\ref{yia}) with the
function%
\begin{equation}
B(u)=\frac{\left( 1+u^{2}\right) ^{-1}-\left( D-1\right) \left(
1-a^{2}/\alpha ^{2}\right) }{2\sqrt{1-a^{2}/\alpha ^{2}}\sqrt{1+u^{2}}},
\label{Bu}
\end{equation}%
and $u=\eta a$. Although, we have obtained the formula (\ref{Bu}) in the
range of parameters assumed for the validity of (\ref{Hypas}), the
corresponding formula for other values of $\eta a$ is obtained by a simple
analytic continuation. Moreover, the result can also be generalized for the
case of AdS space by the replacement $\alpha \rightarrow i\alpha $. Having
the expression for the function $B(u)$, the function $C(u)$ is found from (%
\ref{Cx}):%
\begin{equation}
C(u)=2\frac{C_{k}(x_{a})-1}{\sqrt{1+u^{2}}}\left[ D\left( \xi -\xi
_{D}\right) +\frac{\xi }{C_{k}(x_{a})}-\frac{\left( 1+u^{2}\right) ^{-1}}{%
4C_{k}(x_{a})}\right] .  \label{Cu}
\end{equation}%
With this function, the integrals in the expressions (\ref{phi2Near}) and (%
\ref{T22Near}) of the leading terms in the VEVs of the field squared and the
energy-momentum tensor are expressed in terms of the gamma function.

\end{document}